\pdfoutput=1
\documentclass[11pt,twoside]{article}
\usepackage{asp2014}
\usepackage{mathrsfs}

\def\kms{km~s$^{-1}$}
\def\SiO{SiO~$v$=1, $J$=5-4}
\def\LectureAA{Lecture on Calibration} 
\def\LectureBB{Lecture on Error Recognition} 
\def\LectureCC{Lecture on Imaging} 
\def\LectureDD{Lecture on Advanced Imaging} 
\def\LectureEE{Lecture on Polarization} 
\aspSuppressVolSlug
\resetcounters

\begin{document}

\title{Advanced Gain Calibration Techniques in Radio Interferometry}
\author{Crystal L. Brogan$^1$, Todd R. Hunter$^1$, and Ed B. Fomalont$^{1,2}$} 
\affil{$^1$NRAO, Charlottesville, VA, USA; \email{cbrogan@nrao.edu}}
\affil{$^2$NRAO, Joint ALMA Observatory, Santiago, Chile}

\paperauthor{Crystal~L.~Brogan}{cbrogan@nrao.edu}{0000-0002-6558-7653}{NRAO}{}{Charlottesville}{VA}{22903}{USA}
\paperauthor{Todd~R.~Hunter}{thunter@nrao.edu}{0000-0001-6492-0090}{NRAO}{}{Charlottesville}{VA}{22903}{USA}
\paperauthor{Ed~B.~Fomalont}{efomalon@nrao.edu}{}{NRAO}{}{Charlottesville}{VA}{22903}{USA}

\begin{abstract}

In this lecture, we describe a number of advanced gain calibration
techniques. In particular, self-calibration is an important tool in
interferometric imaging at all wavelengths. It allows the observer to
determine and remove residual phase and amplitude errors that remain
in the data after normal calibration while simultaneously producing a
more accurate and more sensitive image of the target source.  We
describe the basic technique of self-calibration and attempt to dispel
some common misconceptions. We proceed to give a range of useful tips,
and provide continuum, spectral line, and mosaic self-calibration
examples using ALMA data.  We also discuss fast switching and
radiometric phase correction along with advanced phase transfer
techniques that can become advantageous or even essential at high
frequency where the density of sufficiently bright phase calibrators
becomes sparse.

\end{abstract}

\section{Introduction}

In this lecture, we discuss several techniques that go beyond standard antenna based gain (phase and amplitude) calibration (see the \LectureAA). Such techniques are most needed at frequencies for which the atmosphere induces gain fluctuations more rapidly than the cadence upon which it is practical to observe a calibration target. In the radio regime, these fluctuations occur most notably at the lowest frequencies ($<1$~GHz) and highest frequencies (from $\sim 20$~GHz up to $\sim 1$~THz, where the emission becomes opaque to ground-based telescopes), though the principle cause differs. At lower frequencies, gain fluctuations are dominated by variations in the electron content of the ionosphere, while at higher frequencies the fluctuations are dominated by variations in the wet and dry components of the troposphere. Due to the finite angular scale of the variations (as viewed from the ground), the effect of the fluctuations on the data will vary as a function of the baseline length.  Although atmospheric fluctuations affect both phase and amplitude, phase variations are typically the dominant source of error in the data. In what follows we often use the term ``phase'' variations synonymously with gain variations.

\section{Self-Calibration}

\subsection{Definition}


The term self-calibration (hereafter "self-cal") is used to describe the procedure for obtaining a time series of antenna-based complex gains (phase and amplitude) that minimize the deviation of the observed visibilities from a model of the source visibilities. The concept of self-cal forms the basis of radio interferometric calibration.  Indeed, the standard procedure of calibrating interferometric data, by solving for these gains on a point-like calibrator of precisely known position, is simply a special case of self-cal where the model is assumed to have zero phase and equal amplitude on all baselines.   When the morphology of the target is not known {\it a priori}, such as for a science target, one can use an iterative process to gradually construct a better and better image (model) of the target, after standard calibration has been applied.  The technique for doing self-cal on such targets was developed in the 1970's for long baseline interferometry \citep[see the review by][]{Pearson84}, and brought to a high standard at the Very Large Array (VLA) \citep{Schwab80}. Earlier summer school chapters provide a good theoretical description of the process \citep{Cornwell99}. Mathematically, the deviations ($S_k$) of the observed visibilities from the model can be written as:
\begin{equation}
S_k = \sum_k\sum_{i,j}^{i\neq j} w_{i,j} \lvert g_i(t_k)g_j^*(t_k)V_{i,j}^o(t_k) - M_{i,j}(t_k)\rvert ^2
\label{minimize}
\end{equation}
where $i$,$j$ corresponds to the baseline formed between antennas $i$ and $j$, $t_k$ (often called the ``solint'') is the time interval used to solve for the complex solutions: $g_i(t_k)g_j^*(t_k)$, $w_{i,j}$ are the visibility data weights, $V_{i,j}^o(t_k)$ are the complex visibilities, and $M_{i,j}(t_k)$ is the Fourier transform of the input model image. Although it is possible to solve for self-calibration solutions on a baseline by baseline basis, in practice this is highly unconstrained, and indeed generally not recommended unless one is attempting to achieve very high dynamic range and is very experienced \citep[see examples in][]{Perley99,Smirnov11}. Instead, solving for antenna-based self-calibration solutions allows one to take advantage of the highly over-constrained nature of the visibility data from interferometers with significant numbers of antennas. Furthermore, the antenna-based nature of the solutions conserves the phase closure of the data and is essentially equivalent to using closure quantities to model the data \citep{CW81}.  


\subsection{Benefits and Misconceptions}
\label{Miscon}

Self-cal can be used to overcome a wide range of standard calibration
inadequacies. It is a common misconception that self-cal is only
important when atmospheric gain variations happen on shorter
timescales than the switching time between phase-calibrator and
science target. In fact, the transfer of gain solutions between a
calibrator in a particular direction to another direction in the sky
at another time (i.e. your science target) is {\bf never} perfect. The
level of error induced by this lack of perfection depends on many
factors, one of which is of course the weather (or more specifically
the phase and amplitude stability), but it also includes instrumental
limitations.  For example, the accuracy of the antenna positions
limits the accuracy of phase transfer from one position on the sky to
another.  The accuracy scales with the ratio of antenna position error
compared to the observing wavelength ($\epsilon_{pos}/\lambda_{obs}$)
coupled with the angular separation between phase calibrator and
science target \citep[for details see][]{Hunter16}, independent of the weather. For
$\epsilon_{pos}/\lambda_{obs}<$ a few (i.e. not bad enough to preclude
some kind of reasonable initial image model), a single
self-calibration solution per antenna averaged over the entire
observing period can yield substantial improvements in signal-to-noise
(S/N).  Together, instrumental calibration limitations restrict the
achievable imaging dynamic range (S/N) from standard gain transfer to
about 1000:1 and 100:1 for the Karl G. Jansky Very Large Array (VLA)
and the Atacama Large Millimeter/submillimeter Array (ALMA),
respectively (considerably worse for snapshot observations and
typically also worse for higher frequencies). Thus, even if your data
is taken in the most phase-stable conditions, if your time-on-source
would theoretically permit high dynamic range, it will not be realized
without self-cal.

The full range of data corruption effects are described in the
\LectureBB\/, but roughly fall into two categories: effects that are
direction dependent (solutions derived for one direction in the sky
are not correct for a different position) and effects that standard
gain calibration must assume are fixed in time but in fact are time
variable.  As a general statement, when trying to calibrate time
variable effects, the $t_{\rm solint}$ used to solve for self-cal
needs to be short enough to allow for improvement of whatever
deficiencies exist in the science target calibration after the best
standard calibration has been applied, while long enough to ensure
good signal-to-noise on the antenna based-solutions. Additionally, it
is notable that most data corruption effects are a function of the
signal strength, so that the greatest benefits will be seen for data
with stronger emission.

Other common misconceptions about self-cal are that (I) the resulting
image isn't scientifically valid because "you can make an image look
like anything you want with self-cal"; and (II) "If I self-cal, all
the position information will be lost". For modern, many-antenna
interferometers, these statements are untrue, provided that one
follows these simple rules: (a) be sure that there is sufficient
signal-to-noise (S/N) to constrain the model and (b) be conservative
in one's iterative approach. It is true that if you take data that has
no signal, and use a point source model at the phasecenter to
self-calibrate, then you will find that the self-calibrated image
shows a point source at the phasecenter of a few sigma due to
statistical noise peaks in the data being rectified in the image at
the model position! However, the correct response to these concerns is
not to shun self-cal but to never violate rules (a) or (b). Indeed, in
the scenario just described, it is very likely that a large fraction
of the solutions would have failed, and that the solutions vary
randomly and unrealistically with time -- both strong indicators that
all is not well. In the remainder of this section, we provide useful
rules of thumb and some practical examples on the process of
self-calibration. In particular, more details on (a) are described in
\S~\ref{Possible} and more details on misconception (II) and (b) are
given in\S~\ref{imaging}.

\subsection{Procedure}

There are four fundamental skills needed for achieving a successful self-cal: (1) Determining if self-cal is possible for your data; (2) Choosing a good iterative procedure and following it faithfully; (3) Knowing how to optimally image your data; and (4) Learning how to improve the S/N of your solutions. In this section we review these fundamentals.

\subsubsection{Is self-cal possible for my data?}
\label{Possible}

The minimum signal-to-noise required for self-cal solutions depends upon the level of error that you are willing to accept. For example, at a 
particular solution interval ($t_{\rm solint}$), an antenna-based signal-to-noise ratio $(S/N)_{Ant}(t_{\rm solint})\gtrsim 3$ ensures that phase solution uncertainties will be less than $<20\deg$, and $(S/N)_{Ant}(t_{\rm solint})\gtrsim 10$ will induce amplitude solution uncertainties $<10\%$. Henceforth, $(S/N)_{Ant}(t_{\rm solint})$ will be referred to as simply $(S/N)_{self}$. The $t_{\rm solint}$ required to actually substantially improve your data will depend on the dominant sources of error. As described above, errors that are ``directional'' like those induced by antenna position errors, only require a single $t_{\rm solint}$ averaged over 1~hour (i.e. a typical ALMA execution block length) for correction. Errors that are time dependent require $t_{\rm solint}$ shorter than the time for significant change to occur. 

An easy way to determine $(S/N)_{self}$ is to make an initial image using the best practices for your observing mode (see \S~\ref{imaging}) below. Measure the rms noise in a representative signal-free region of the image. Convert this noise, $\sigma_{image}$, to that appropriate for a single antenna using the formula: $\sigma_{Ant}=\sigma_{image}\times \sqrt{n-3}$ where n is the number of antennas \citep{Cornwell81}. Then for a given $t_{\rm solint}$, 
\begin{equation}
\sigma_{self}=\sigma_{image}\times \sqrt{n-3}\times \sqrt{\frac{t_{\rm on\_source}}{t_{\rm solint}}}.
\end{equation} In the simplest and most conservative assumption, the ``signal'' is the peak intensity ($I_{peak}$) in the image, so that $(S/N)_{self}=I_{peak}/\sigma_{self}$. However, it should be emphasized that $\sigma_{image}$ will improve with self-cal, thus one should in principle re-evaluate the $(S/N)_{self}$ at each stage of the process. Additionally, there are a number of caveats that add complexity to this simple calculation. 

The trickiest question, which in fact leads to almost all of the caveats, is what is the relevant ``signal'' for self-cal? For modern, many element interferometers, each individual antenna will have data from a range of baseline lengths. Depending on where it is in the array, a particular antenna can have mostly short, mostly long, or mostly intermediate baseline lengths. This means that for a complex science target morphology that includes extended emission, all antennas do not ``see'' the same signal due to spatial filtering. Antennas with mostly short baselines will have as their signal something approaching the integrated flux density (rather than just the peak intensity), while antennas with mostly long baselines will (mostly) only be sensitive to the peak or point-source emission. This underlying complexity makes a realistic determination of $(S/N)_{self}$ antenna dependent, and closely coupled to the often unknown ({\it a priori}) exact source distribution. The bottom line is not to get impeded by the details of the exact calculation but to gain a general sense of how your baselines are distributed compared to how your source is distributed and what that is likely to imply about the actual $(S/N)_{self}$ per antenna, especially the ones dominated by long baseline lengths. 

\begin{center}
\vspace{-2.5mm}
\fbox{\begin{minipage}{0.95\textwidth}
As a general rule of thumb for an array with 25 antennas, if the $(S/N)_{image}>20$ it is usually worth at least attempting phase-only self-calibration on a time-scale of several minutes. This rule of thumb scales approximately with the number of antennas.
\end{minipage}}
\end{center}

\noindent However, as with all scientific endeavors it is up to the scientist to ensure that the results are valid and reliable. If after your best efforts (and usually much trial and error), the results do not seem like an improvement, simply don't use the self-cal image -- it is still likely that you will have learned much about your data during the process. After standard calibrator based calibration, significant deviations of the self-cal phase solutions from zero or the amplitude gains from one indicate deficiencies in either the standard calibration and/or with data flagging. The self-cal solutions, will for example, highlight periods of time in which the phases are particularly badly decorrelated or an antenna is misbehaving due to a hardware issue (such as low gains). Flagging such highlighted time ranges/antennas can bring about a significant improvement in your image even if you do not chose to apply the self-calibration itself.

\subsubsection{Basic Procedure}
\label{Basic}

Self-cal almost always starts with phase-only solves, with amplitude solves only attempted after the phases are refined,
for the following three reasons: (A) antenna-based phase solutions are more constrained than antenna-based amplitude solutions, (B) the largest improvements will typically come from reducing the phase error, and (C) phase solutions are less sensitive to uv-coverage based image artifacts etc.,  In general, the success of the process is controlled by the signal-to-noise available to the antenna-based self-calibration solve $(S/N)_{self}$. Thus, the basic procedure of self-calibration is to 
\begin{enumerate}
\item Split off your science target data after applying all of the standard antenna based gain tables.
\item Back up the current flag state of the data. 
\item Make an initial image, the Fourier transform of which is stored as a model.
\item Do a phase-only self-calibration on a medium to long time-scale using that model (see more on this choice below).
\item Apply solutions to the original data.
\item Re-image. If improved (judged by assessing both the change in the peak emission and the rms noise), then repeat the process with shorter and shorter $t_{\rm solint}$ until either the $(S/N)_{self}$ drops too low for good solutions to be achieved, the $t_{\rm solint}$ is as small as the visibility integration time, or no further improvement is seen. Note it is always important to assess the image S/N, not just the noise when assessing the level of improvement.
\item Attempt to solve for amplitude self-calibration solutions, typically with a longer timescale, while applying the best phase solutions obtained so far.
\item If significant improvement is seen, the amplitude self-calibration can also be performed using shorter $t_{\rm solint}$, but typically this will show lower levels of improvement as amplitude (both instrumental and atmospheric) tends to intrinsically vary more slowly than phase.
\end{enumerate}

\subsubsection{Understanding the Mechanics of the Software}

Self-cal is always iterative. Thus, one needs to make a choice about the fundamental nature of the iterative approach to be used and follow it faithfully. Some of the confusion that arises about these approaches is that the exact procedure is rather dependent on the software one is using. In this Lecture, we assume the use of CASA \citep{McMullin07} in which the visibility data resides in measurement sets containing one or more columns.  The original data resides in the DATA column and if you run {\tt applycal} the corrected data resides in the CORRECTED column; each time {\tt applycal} is run the new calibration overwrites the previous CORRECTED column. It is important to recognize that, in CASA, the solver {\tt gaincal} looks at the DATA column and the imaging code {\tt clean} looks at the CORRECTED column (though a new task in CASA 4.5 called {\tt tclean} allows one to select the column for cleaning). Also, the solver {\tt gaincal} creates a calibration table that you can chose to apply or not using {\tt applycal}. In all cases, the self-calibration should start at the point where the complete suite of standard calibrator-based, antenna-based calibration has been applied to the science target and the resulting CORRECTED column has been split off such that in the new dataset, the DATA column contains fully calibrated data. A final feature of {\tt applycal} that is very important to take into account is that it will {\bf flag} the data that correspond to the timeranges of the failed solutions in your self-cal (unless the task is told otherwise which is not generally recommended). For the most part, a large number of failed solutions is a strong indication that something is going wrong -- either there is inadequate $(S/N)_{self}$ for your chosen $t_{\rm solint}$ or the model is inadequate, improperly stored, etc. For this reason, it is important to back up the flag state of your data periodically so that it is easy to recover from mistakes and start again.  

In order to produce a successful iterative self-cal, you must understand how to manage and apply the solutions that you obtain. There are three distinct approaches to managing iterative solutions. Depending on the software you use, some approaches are more practical that others.  In the approach that we will call {\bf serial}, one always uses the best current image model to do the phase-only solves on the original data (i.e. in the original DATA column). To obtain the final phase-only calibration, one only applies the final phase-only table. For the second approach, which we call {\bf incremental}, one applies each previous phase-only solution on-the-fly while solving for the new solutions. The resulting tables are incremental, and one must apply each of them to get a complete correction, but one still goes back to the original data for each solve. The final distinct approach, which we call {\bf incremental+split} one applies each previous phase-only solution and splits off the CORRECTED column so that in the new dataset the corrected data is in the DATA column. In that case new tables are incremental but one also generates a new corrected dataset each iteration, and at the end, one only applies the final table to the final split dataset. This approach is definitely the most expensive in terms of bookkeeping in CASA.  If done consistently, any of the approaches can work in most situations. However, especially for data near the threshold of being self-calibratable, both of the incremental phase-only methods carry the danger of ``locking-in'' poor properties of early rounds of self-cal when the model is inevitably not as good as it is in later rounds. The examples in this Lecture use the {\bf serial} method for phase-only self-cal and the {\bf incremental} method for subsequent amplitude self-cal. 

\subsubsection{Imaging for Self-calibration}
\label{imaging}
Self-cal fundamentally depends upon the quality of the input image model, which in this context corresponds to the Fourier transform of the clean components. The more closely the image model correctly represents the data, the better the self-cal results can be. The many vagaries of interferometric imaging are covered in the \LectureCC\/ and the \LectureDD\/. We wish to emphasize only a few points here: (1) if specialized imaging techniques are needed to achieve high dynamic range imaging given your observing mode (widefield, wideband, mosaic, polarization, etc.) those techniques need to be used for the imaging throughout the self-cal process as well; (2) one must provide constraints on the clean process to limit the clean components to those that are very trustworthy at the beginning stages and only open up to less certain emission (or absorption) as the process goes on. This is best accomplished presently through interactive cleaning at each stage and the use of clean masks to limit the cleaning to high confidence regions. Additionally, if the uv-coverage limits the imaging (e.g. too little data at short baselines, a gap etc.), one can experiment with uv-ranges or tapering to improve the imaging. Generally, if the imaging uses a limited portion of the data, then the self-cal solve should be similarly limited. 

If there are no high confidence areas in your initial image, you probably ought not to be doing self-cal. There are of course exceptions such as VLBI where it is not uncommon to to start with a point source model at the phasecenter, ignoring the intrinsic position and initial morphology information. This method should only be attempted when the uv-data reveal the presence of strong signal that is simply too decorrelated to make a reasonable first image. When this technique is employed, the absolute position information of the image is lost and is at the heart of misconception (II) described in \S~\ref{Miscon}. However, for connected element interferometers taken under reasonable weather conditions for the observing frequency and a well-performing instrument, it is rarely the case that a usable starting model cannot be obtained from the initial image provided there is enough S/N. In such a case the images from subsequent self-cal iterations will have their positions intrinsically tied to the initial image, which in turn is tied to the phase calibrator position, but there is some freedom for the positions to shift by small amounts. Take for example, a simple case of a fairly strong source that is intrinsically unresolved, but is smeared out initially due to poor phase coherence. Self-calibration will make the emission correctly point-like, but the peak of emission may not be precisely at the location it started (rather it will be at the pixel closest to the centroid of the initial smeared out structure). Therefore, it is helpful to over-sample the synthesized beam by a factor of $\sim 5$ in the smallest dimension. Provided that the S/N was high enough to be constraining in the first place, typical shifts will be less than one image pixel. Nevertheless, it is always a good idea to check the absolute position integrity of positions before and after self-cal (see for example \S~\ref{MiraCont}). 

\begin{center}
\vspace{-3.mm}
\fbox{\begin{minipage}{0.95\textwidth}
Additionally, keep in mind that while you cannot make real but weak emission vanish by NOT including it in a clean mask/self-cal model, you can however, create weak features by including noise or image artifacts in your self-cal image model. Thus, be conservative in the initial stages!
\end{minipage}}
\end{center}
\noindent Especially if you are particularly interested in a weak source in a field of strong emission, do not include it in your clean model until either near the end, or even at all (assuming it contributes very little to the integrated total flux). Indeed, not including such emission in the model/self-cal is an excellent way to check the reality of weak emission.  If it is real emission, then its S/N should improve, while emission due to calibration artifacts and sidelobes should diminish as the self-cal process progresses.

\subsubsection{Sixteen Tips for Improving the Solutions and Results}
\label{Tips}

Generally speaking, self-cal solutions can be derived for the continuum and applied to your spectral line data, and vice-versa. One should use the strongest signal/noise data available to do the self-cal. In the fortuitous situation that you have both strong continuum and line emission, the very best result will often be obtained from applying the self-cal solutions derived from itself. This may seem strange -- shouldn't temporal correction apply to all the data taken over those times? Yes, but those aren't the only effects that get subsumed into the solutions. For example, self-cal carried out on a narrow spectral line feature is very insensitive to delay like errors (errors that change as a function of frequency), while modern wide-band continuum capabilities lead to continuum self-calibration solutions that include delay-like contributions. 

In the case that a significant number of failed solutions are being reported, it is likely that you do not have adequate $(S/N)_{self}$ with the current inputs. In addition to the obvious measure of increasing $t_{\rm solint}$, there are a number of other parameter adjustments and actions that can be used to try to increase the number of successful solutions and/or the overall results. 
\begin{enumerate}
\item First ensure that your data have proper visibility weights ($w_{i,j}$) $\propto 1/(T_{sys\_i}T_{sys\_j)}$ (where $T_{sys_i}$ is the system temperature for antenna i), at minimum, and preferably also including modification by the antenna gains ($g_i^2g_j^2$), using {\tt calwt=True} when applying the phase calibrator's amplitude gains \footnote{See the following guide for ALMA data weights in CASA: \url{https://casaguides.nrao.edu/index.php/DataWeightsAndCombination}}.  If accurate visibility weights are not available, or suspect for any reason, the CASA task {\em statwt} can be used to derive reliable $w_{i,j}$ (just be sure to exclude line emission from the calculation). The self-cal solve takes these weights into account in order to up-weight the highest sensitivity data (see Eq.~\ref{minimize}). 
\item In addition to inspecting the temporal behavior of the solutions with {\tt plotcal}, use the {\tt yaxis='snr'} option to investigate the S/N of the solutions. This will give important insight as to the reliability of the solutions, and can highlight particular antennas or groups of antennas (long baseline ones for example) that have less significant solutions. If for example, one antenna stands out as having much lower S/N solutions than the rest, you might try flagging that antenna.  If it is a group of antennas, or all antennas, you may try one of the other approaches below to increase the $(S/N)_{self}$.
\item Check whether the model compares well with the data.  Make a plot of the Fourier transform of the image model in the form of amplitude vs. uv-distance (for example using {\tt plotms} to plot the MODEL) and compare it to a similar plot of the data. Are the two comparable? Is there any model (it is possible to turn off the saving of the model in {\tt tclean} for example)? Try limiting  the solve to the portion of the uv-plane where the model and data are well matched. Does this improve the solutions/result? Note, excluding things other than antennas from the solve does not necessarily lead to {\tt applycal} flagging those things -- unless the exclusion results in there being no antenna-based solutions. Obviously, if for example the uv-range becomes too restrictive, then you can lose whole antennas from the solve entirely and then they will be flagged by {\tt applycal}.
\item Related to the previous item, in particular for amplitude solves, is your model missing flux at short baselines compared to the data?  This issue can be tricky, because if the model is missing flux and you do nothing to ameliorate the mis-match, the resulting rms noise will often go down and the image may even "appear" better. However, on closer inspection the peak intensity and flux density will have also decreased. If the model is missing only a modest amount of flux one can try the {\tt solnorm=True} option. This will keep the ensemble amplitude correction around 1, and will generally prevent an overall reduction in the integrated flux. However, it also limits the freedom of the self-cal to remove errors, and thus its ability to match the model. Alternatively, one can exclude the shorter spacings from the solve with an inner uv-range cutoff. Provided that the cut is not so severe as to preclude antenna-based solutions for inner antennas, this often works the best from the point of view of preserving the flux and still allows for antennas needing largish deviations to get corrected. It is often advantageous to try both ameliorating techniques for the amplitude solve: an inner uv-range cutoff versus the {\tt solnorm=True} option and see which gives the most improvement in the image S/N without a loss in integrated flux. 
\item A good rule of thumb for amplitude solves is that for data taken from a well performing array, in appropriate weather conditions, amplitude solutions deviating by more than about $20\%$ from 1.0 are suspect and one should investigate the underlying cause. Another useful option for amplitude solves, is to perform a {\tt calmode='ap'} solve rather than just {\tt 'a'}. Since one applies the best phase-only solution on-the-fly at this stage, significant deviations (larger than a few degrees) of the phases from zero in the {\tt 'ap'} solve is an indication that the phases were not in fact sufficiently refined in the previous phase solve and/or the amplitude solutions are not well constrained. 
\item When self-cal is performed on a target with extended emission, it is very important to assess not just the change of peak emission compared to the rms noise, but also how the integrated flux density and rms noise (and their ratio) compare with previous iterations. Even a small improvement in the peak emission can equate to a large improvement in the integrated flux density for extended emission. Be sure to measure the integrated flux density consistently when doing the comparison.
\item Choose a good reference antenna ({\tt refant}), typically a well-behaved antenna from near the center of the array. CASA will switch to an alternate reference antenna when the selected one has a failed solution (this can be due to previous manual or on-line flagging, or $(S/N)_{self}$ less than the minimum required). This change can make interpretation of the solutions challenging if it happens often. Additionally, a significant number of "using alternate refant" warnings indicates that the chosen {\tt refant} is not "well-behaving" and you should try a different one. If a change to a better behaving {\tt refant} still results in lots of "using alternate refant" warnings, then this often indicates that $t_{\rm solint}$ is too short.
\item Try averaging the parallel-hand polarization products before solving by using a {\tt gaintype='T'} solution rather than the default {\tt 'G'} solution (i.e. XX and YY for ALMA; RR and LL for JVLA), for a $\sqrt{2}$ improvement in $(S/N)_{self}$. In principle, the instrumental phase offset between the parallel hand products will have already been removed by the standard bandpass calibration, provided they are stable in time. However, this is only a good approximation if the original phase calibrator and the science target are unpolarized, or a full polarization calibration has also been performed and applied. Otherwise, especially if you want to apply the solutions to other parts of the data with different polarization characteristics, it is safest to use the average. Even if using the {\tt 'T'} solutions to apply to the data, it can be useful to make a {\tt 'G'} type solution and investigate its properties with the  {\tt poln='/'} option of {\tt plotcal}. This plot (for a {\tt 'G'} solution) will show the ratio of the two parallel hands, and for unpolarized data is a sensitive indicator of the reliability of the solutions as (non-instrumental) temporal variations should track in the same way leading to a ratio near one. Random variations in the ratio indicates either a significant (uncalibrated) instrumental problem or low $(S/N)_{self}$. In contrast, for data from a polarized target, if polarization calibration has not been performed, the ratio will show smoothly varying (but not random) temporal variations due to rotation of the parallactic angle. 
\item For self-cal on the continuum, try combining spectral windows (observed simultaneously) using the {\tt combine='spw'} parameter in the solve. This can produce a large improvement in $(S/N)_{self}$, as much as $\sqrt{4}$ for ALMA Bands 3-8 and $\sqrt{8}$ for Bands 9-10 (when 90$^\circ$ Walsh switching is employed and the image sideband is stored), and up to $\sqrt{64}$ for JVLA for typical continuum setups. Standard calibration will have removed the instrumental spectral window to spectral window offsets (provided they are stable in time). One does have to be careful however, not to average over significant spectral index variations without taking that into account in the image model (see \LectureDD\/). Also, be cautious about including spectral windows in the self-calibration that have strong atmospheric lines, as they will have lower S/N and in the case of high frequencies can have extra dispersive wet and dry phase contributions compared to cleaner parts of the spectrum (\S~\ref{btob}). Additionally, when combining over wide bandwidths, it is very important to have ensured very good calibration of delay-like errors in the data through the bandpass calibration.  
\item For self-cal on spectral lines, try using an average of all the strongest channels in both the image model creation and the solves, rather than just the peak channel. As long as the data that goes into creating the image matches the data being used for the solve, the temporal variations of the solutions will be valid for each channel individually. The self-cal solutions thus obtained can later be applied to each channel. It can be convenient to split off this group of channels into its own dataset for ease of bookkeeping.
\item For self-cal on mosaics, only use fields in the solve that have strong signal. This can be assessed by plotting the clean model of each field using {\tt plotms}. If the time between visits to the same mosaic field is long (i.e. the mosaic is large), and one needs to use a $t_{\rm solint}$ longer than a single visit to get adequate $(S/N)_{self}$, one can consider combining fields in the solve. 
\item In CASA, the default $(S/N)_{self}$ required for each solution is 3 (i.e. as mentioned in \S~\ref{Possible}) because this ensures errors less than $20^\circ$ for phase-only solves. For data with large numbers of antennas (>25) and hence lots of redundancy, and at least some high S/N solutions (typically on antennas with mostly short baselines), it can be useful to lower the minimum $(S/N)_{self}$ to $\sim 2$ in order to salvage the longest baseline antennas (which for an extended source ``see'' much less of the total flux). It is not recommended to reduce the $(S/N)_{self}$ requirement for data that is near the threshold of being viable for self-cal on most antennas, nor for amplitude solves because they are less constrained. A caveat to this recommendation is in the case that the required corrections are unusually large for most antennas -- then one can accept a greater uncertainty on the correction and still do more good than harm. However, such large corrections are rarely necessary for a well-performing array. 
\item The default number of baselines that must be present to attempt a solve is 4 (though as few as two baselines is technically possible since most modern software does not use closure quantities to constrain the solve). It is not recommended to drop the number of required baselines below 3 and indeed, some prefer raising it to 6, especially for amplitude solves.
\item When combining data from multiple executions, whether they be from the same configuration, different configurations, or the same configuration but different observing cycles, it is a good idea to make individual images for each dataset first to assess whether there are any significant differences in image quality and/or in source position (that cannot simply be attributed to a change in angular resolution). Differences in position could be due to, for example, proper motion, refinements to the phase calibrator position, or if different phase calibrators were used, errors in one or more of those positions. Significant position changes should be reconciled before combining the data and proceeding with a joint self-cal. If you are more confident in one execution's/configuration's positions than another, using the image model from that dataset as the starting model for the other ones will align all the data and the resulting images. Likewise, for executions of variable quality from a phase coherence/weather point of view, it can be advantageous to use only the best one(s) as the initial model for all of them. Additionally, after combining the data, the {\tt applycal} option {\tt interp='linearperobs'} can be used to prevent interpolation of per execution solutions across executions. 
\item When applying solutions created from averaging spws, or just from one spw to another, it is important to specify in the {\tt applycal} exactly how the spws labeled in the solutions table are mapped to the spws you want to apply solutions to using the {\tt spwmap} parameter. For example, for a dataset with four spws (0, 1, 2, and 3), and a {\tt gaincal} run with {\tt combine='spw', spw='1,2'}, when applying that solution to the full dataset you will need {\tt spwmap=[1,1,1,1]}. There are four entries because there are four spws needing correction, and the number is 1 because the spw name in the solution table will be the first spw number used in the solve.
\item If you decide to start over, be sure to restore the data to a fully pre-self-cal state with {\tt clearcal} and {\tt delmod}. Additionally, because the {\tt applycal} steps may have flagged data (i.e. if there have been failed solutions), it is essential to restore the flag state to the one you started with (this is the reason an explicit backup of the flag state is recommended before beginning self-cal (see \S~\ref{Basic}).

\end{enumerate}

\subsection{Self-calibration Examples}

In the next three sections, we describe in some detail the
self-calibration process for a representative strong continuum and
spectral line ALMA use case, as well as a comparatively weak continuum
mosaic ALMA use case. However, the procedures are by no means
exclusive to ALMA and would apply to similar use cases for the VLA,
NOrthern Extended Millimeter Array (NOEMA), Submillimeter Array (SMA),
etc. In order to facilitate giving specific information, the examples
are presented in terms of the corresponding CASA parameters. However,
much of the process should be translatable to other packages.

\subsubsection{ALMA Mira Continuum}
\label{MiraCont}

For this self-calibration example we have chosen the ALMA Science
Verification data at 1.3~mm (Band 6) for the Asymptotic Giant Branch
star Mira taken during the 2014 ALMA Long Baseline Campaign\footnote{
These data and additional information are publicly available
from \url{https://almascience.org/alma-data/science-verification}} \citep{Fomalont15}. The
baseline lengths in these data range from 15.2~m to 15.2~km, and given
the long baselines, one expects that there will be significant
time-variable phases despite the application of the water vapor
radiometer corrections (see \S~\ref{wvrs}) and relatively fast
switching (see \S~\ref{FS}). Each Mira scan in these data is only
$\approx78$~seconds long, interleaved with 21~second phase calibrator
scans; the visibility integration time is 3.02~seconds. These values
are important observational pieces of information to gather because
common choices for the self-cal $t_{\rm solint}$ range from the scan
length on the science target down to the visibility integration time
(or an integer number of them). Two executions make up the total
dataset, both of which have one wide 2~GHz spectral window (spw) for
the continuum and five narrow spws centered on various lines of
interest. When the two executions are concatenated, the total number
of spws doubles because spws with slightly different observing
frequencies will remain distinct. They have slightly different center
frequencies because the exact Doppler shift required to center the
lines is calculated at the start of each execution.

\begin{figure}[h!]    
\centering
\includegraphics[width=1.0\textwidth]{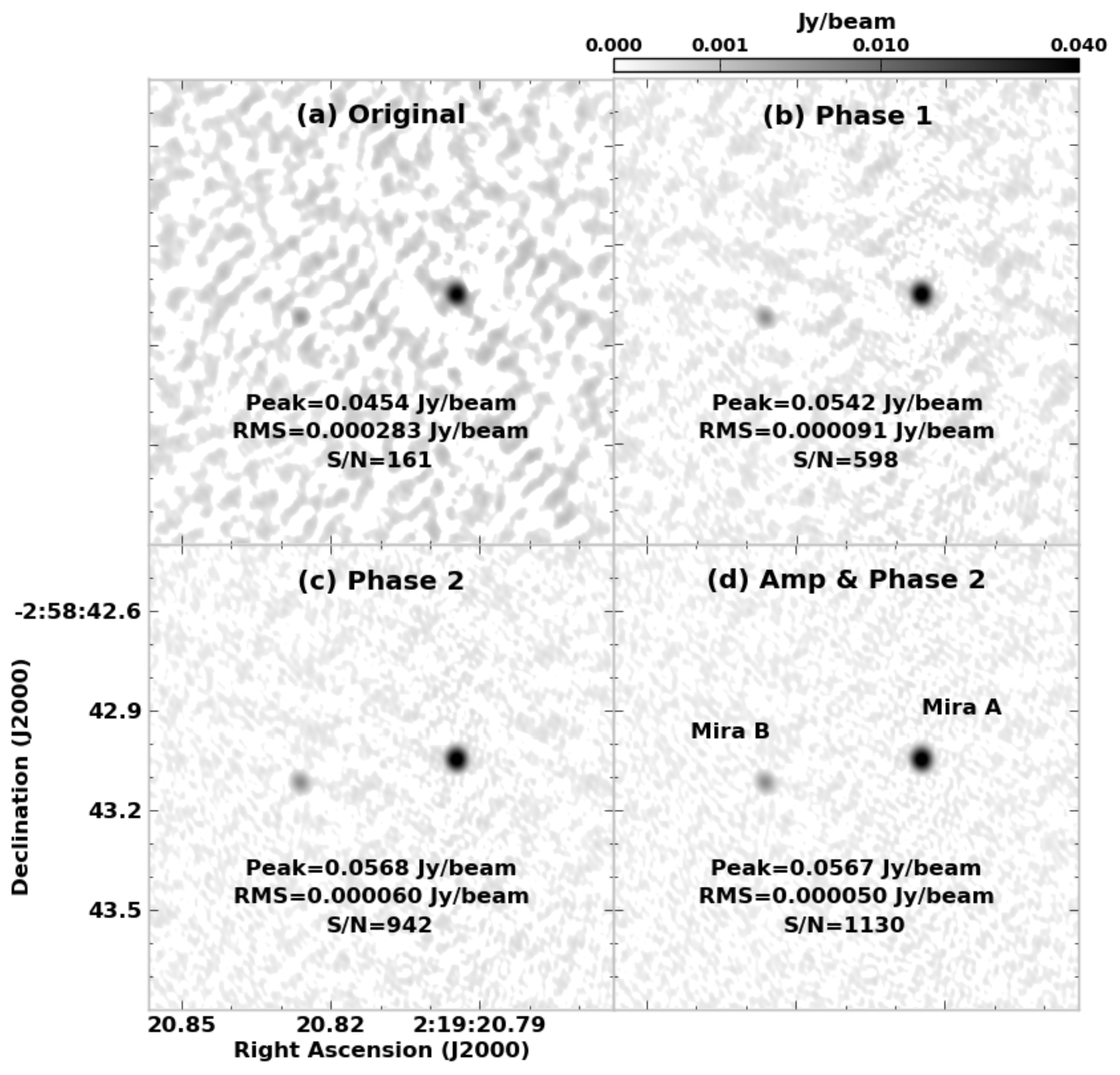}
\caption{Panels show a sequence of images with indicated self-calibration for ALMA continuum data at 1.3~mm for the evolved star Mira. Note that each panel shows the same logarithmic grey scale.}
\label{MiraContself}
\end{figure}
 
\begin{figure}[h!]   
\centering
\includegraphics[width=0.48\textwidth]{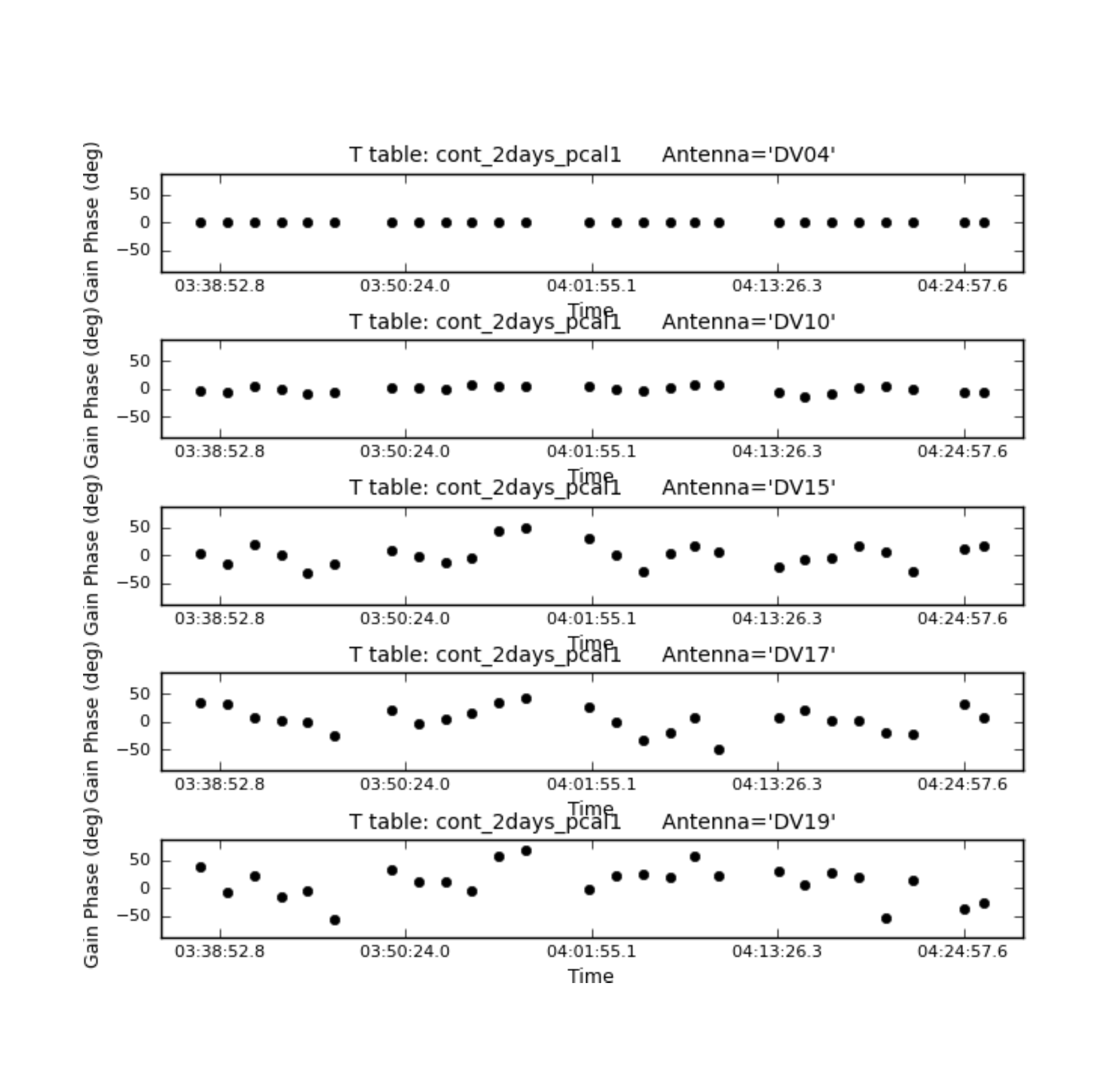}
\includegraphics[width=0.48\textwidth]{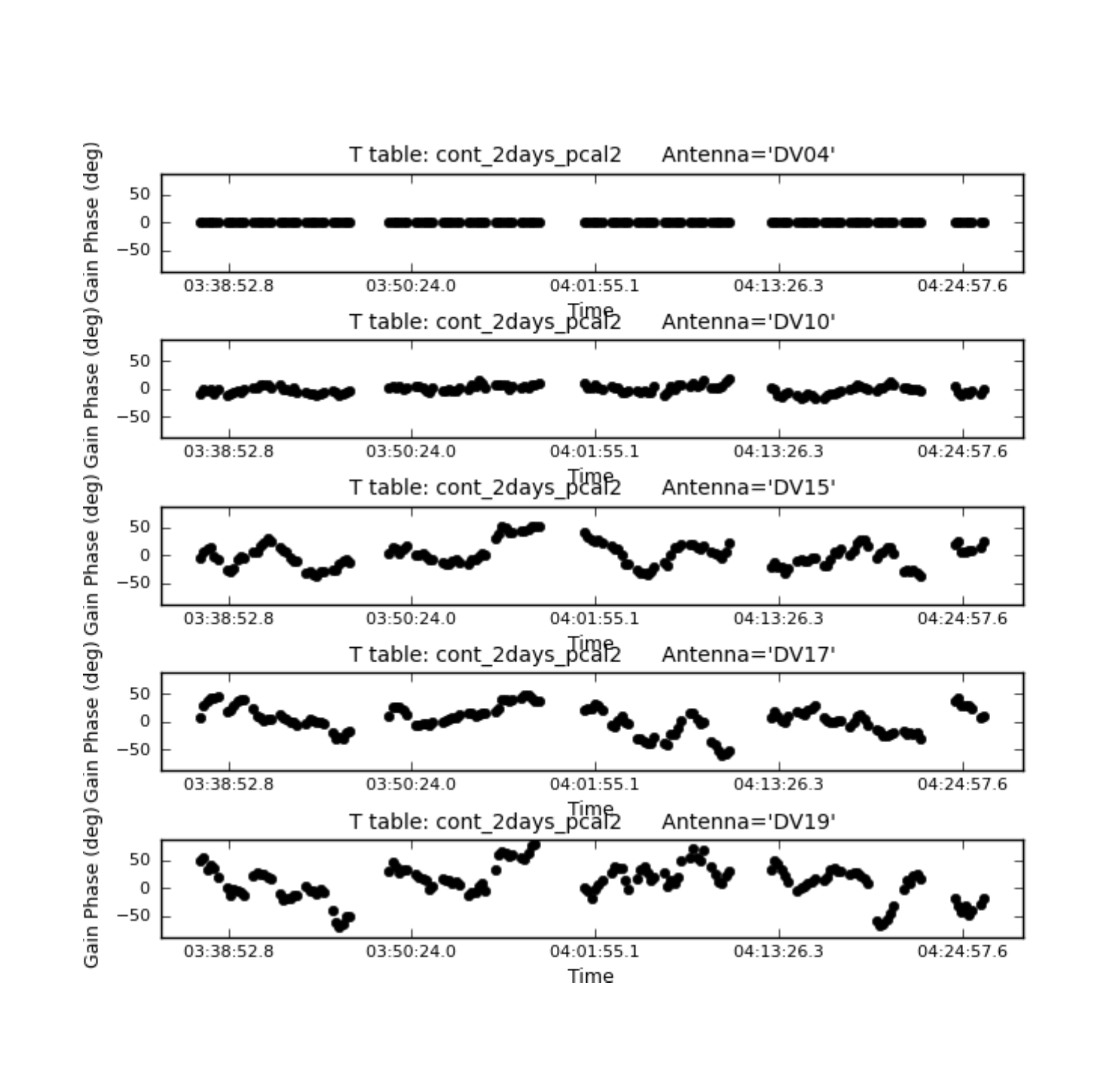}
\caption{Representative phase vs. time plots of the Mira continuum phase-only self-cal solutions for the second execution. LEFT: First phase-only iteration with $t_{\rm solint}$=scan length. RIGHT: Second phase-only iteration with $t_{\rm solint}$=12 seconds. The plotted antennas are ordered with the reference antenna in the top row and antennas located increasingly further from the array center in subsequent rows.}
\label{MiraContPhase}
\end{figure}

\begin{figure}[h!]   
\centering
\includegraphics[width=0.85\textwidth]{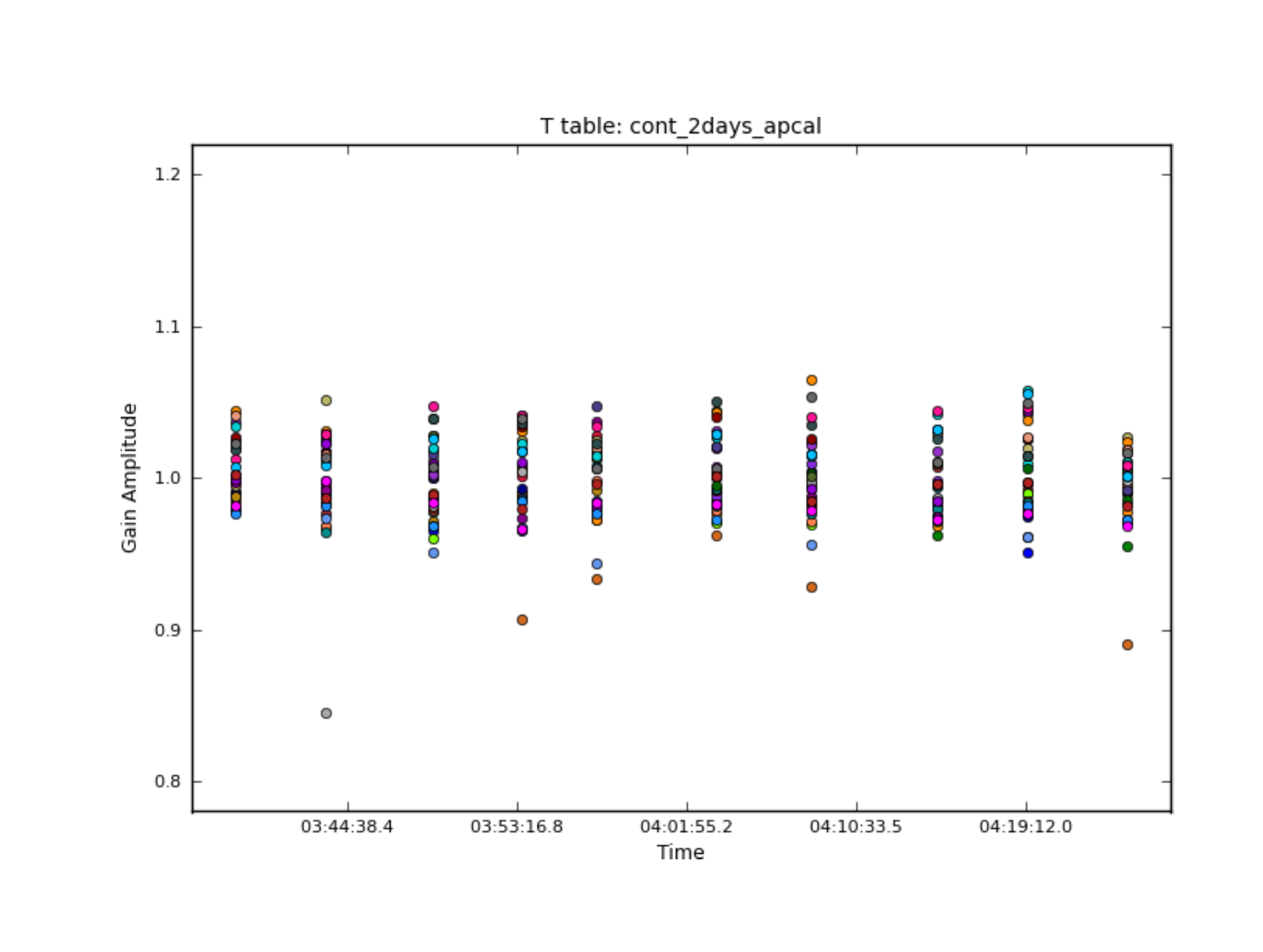}
\caption{Amplitude vs. time plot of the Mira continuum amplitude self-cal solutions for the second execution (points are colored by antenna). The second phase-only solution was applied on-the-fly before solving for these solutions.}
\label{MiraContAmp}
\end{figure}

\begin{figure}[h!]   
\centering
\includegraphics[width=0.45\textwidth]{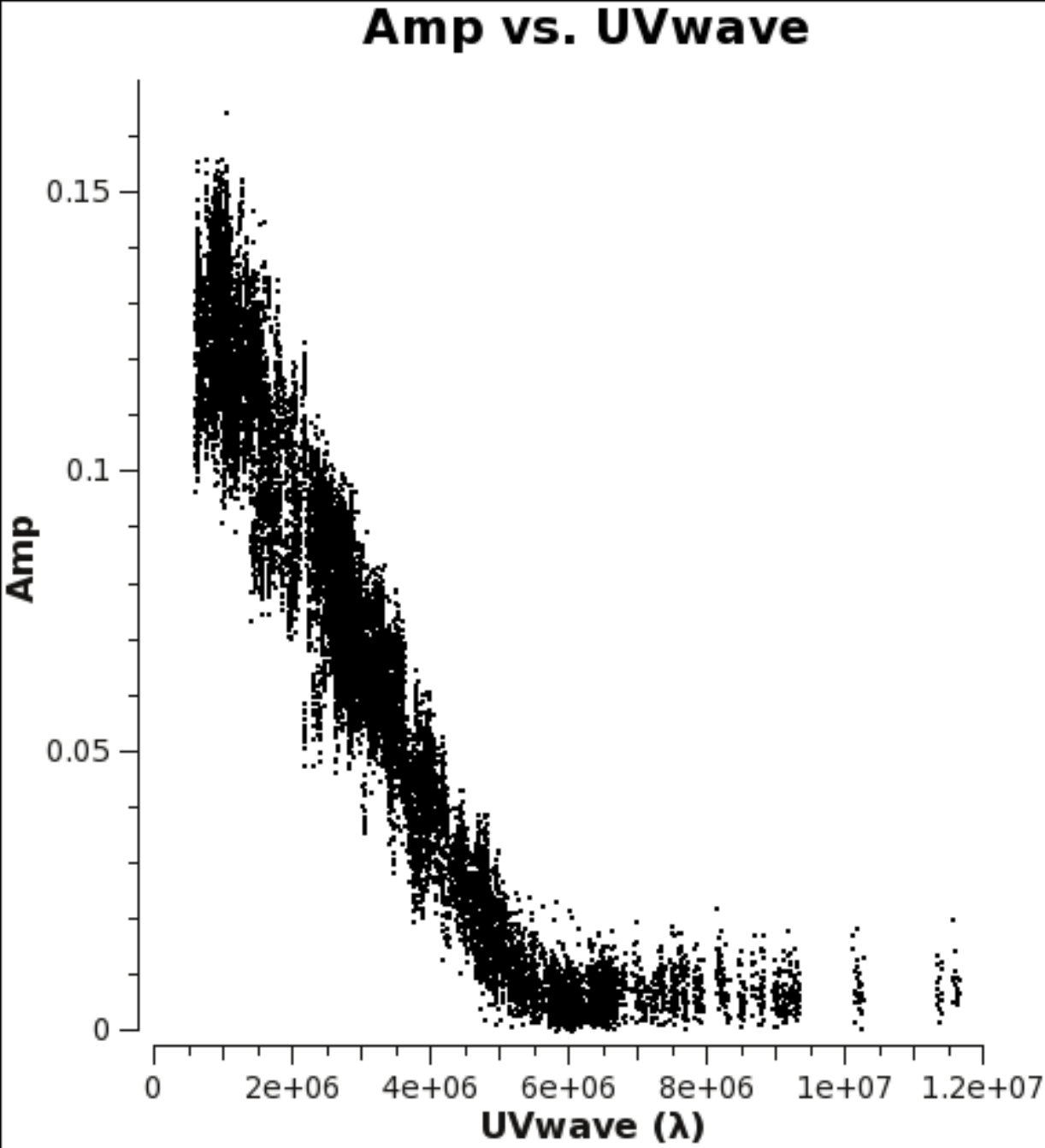}
\includegraphics[width=0.45\textwidth]{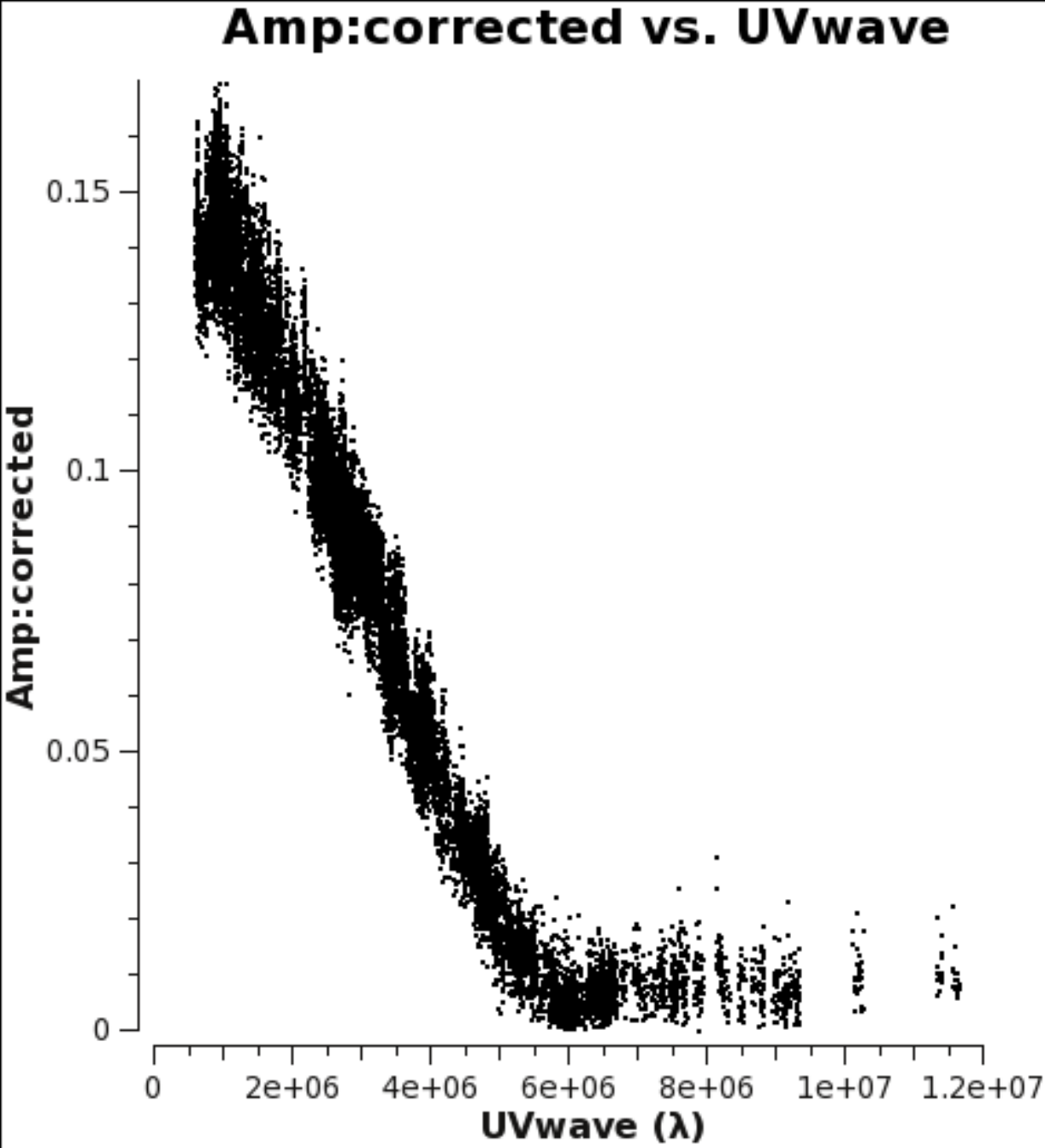}
\includegraphics[width=0.45\textwidth]{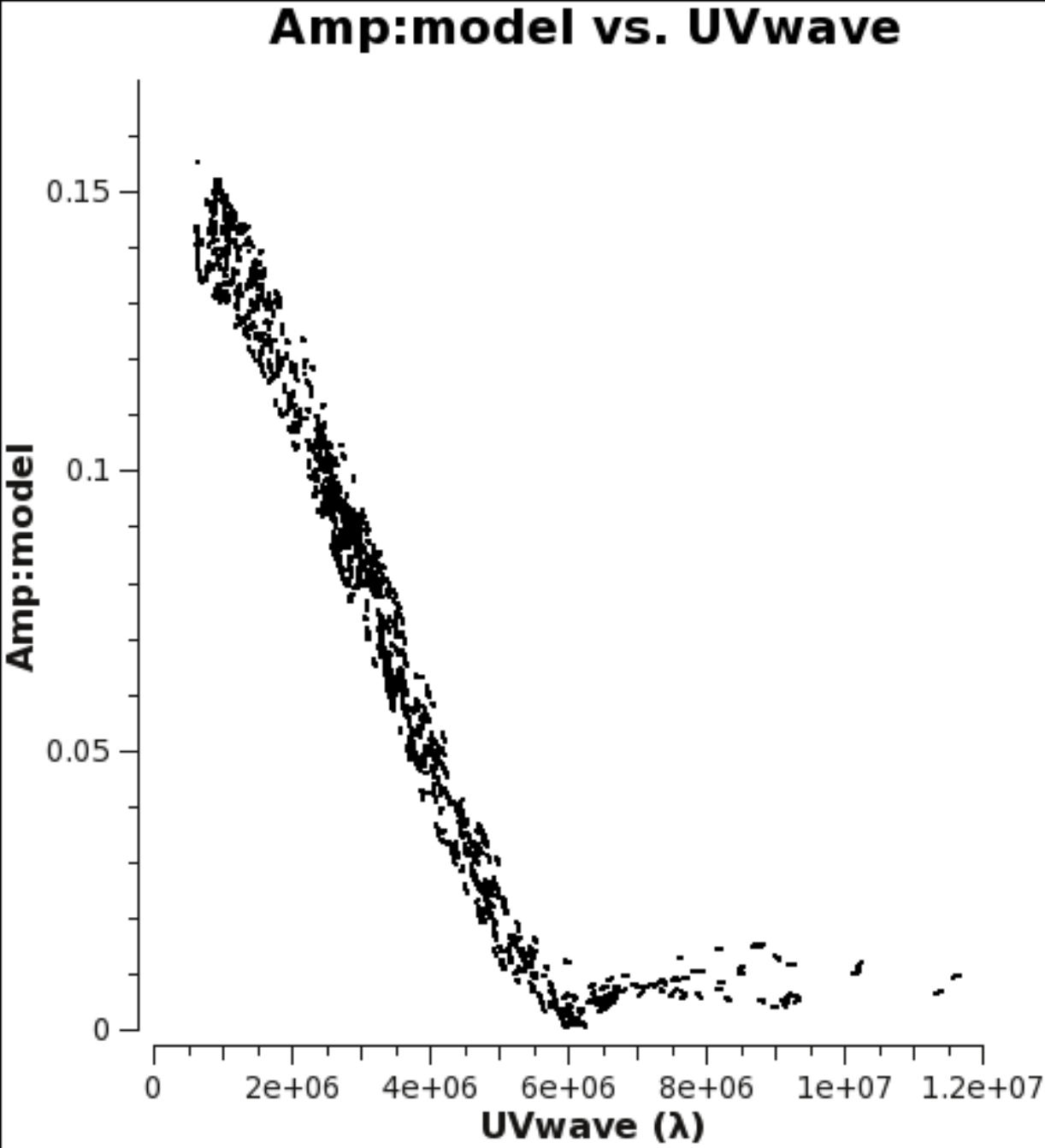}
\includegraphics[width=0.45\textwidth]{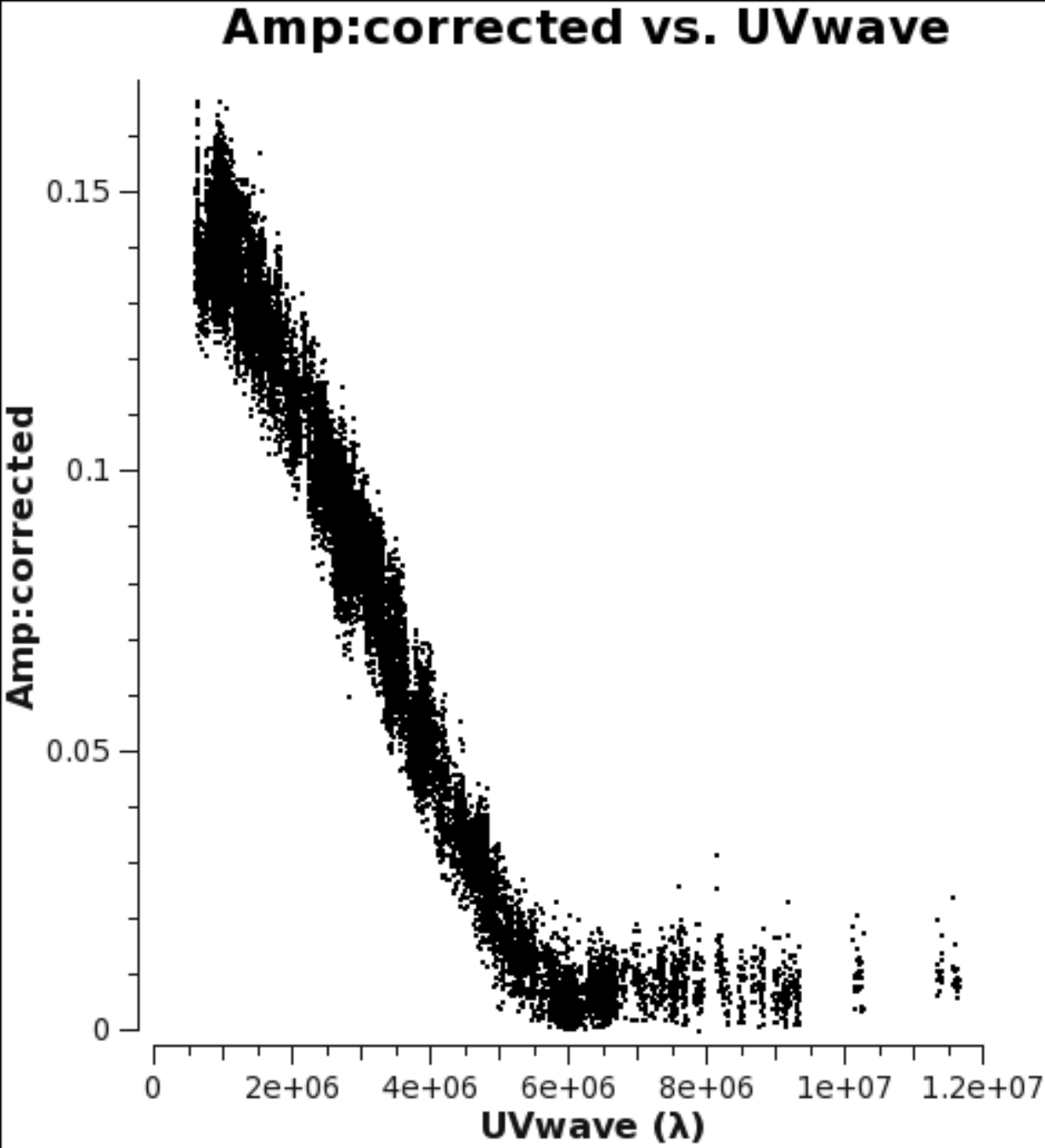}
\caption{Plots of Mira continuum uv amplitude versus wavenumber for (TOP LEFT:) Original data, (TOP RIGHT:) Data calibrated with second phase-only self cal, (BOTTOM LEFT:) The second phase-only Fourier transform of the image model, and (BOTTOM RIGHT:) Final corrected data with the amplitude and second phase corrections applied.}
\label{MiraContUV}
\end{figure}

To begin the continuum self-calibration, we first made a multi-frequency synthesis (mfs) image, putting a relatively tight circular clean mask on the dominant source (Mira A) and cleaning 100 iterations (see Fig~\ref{MiraContself}a). The peak, rms noise level, and S/N are printed on the figure. Note that if we were not doing self-cal but instead making a final image for analysis we certainly would have tried to clean more deeply until the residuals were more noise-like. In this case, however, with such a bright source to start off the self-cal we can be quite conservative -- indeed, the first image has a S/N of 161, consistent with expectations from a dynamic range limited basic calibration, but well below what the ALMA sensitivity calculator \citep{Bridger12} would have predicted for the rms noise based solely on the radiometer equation \citep[see the Lecture on Antennas and Receivers,][]{HunterNapier16}. We then ran the first iteration of phase-only self-cal with a {\tt solint='inf'} and {\tt gaintype='T'} to average the XX and YY polarizations before solving, and {\tt minblperant=6} (everything else was left at its default values). The use of {\tt solint='inf'} and {\tt combine=''} (the default), is shorthand for solving for one solution per scan, and is often a good starting point for relatively high S/N data. If a $t_{\rm solint}$ larger than the scan length is desired, in addition to setting the $t_{\rm solint}$ larger than the scan length (for example {\tt solint='10min'}), you must also set {\tt combine='scan'}. One solution for the entire time can be easily achieved with {\tt solint='inf'} and {\tt combine='scan'}. The two continuum spws were not combined because they come from different execution/times and thus on the timescale of a single scan there would be nothing to combine. If there had been multiple spws per execution that we wanted to use for the continuum, we would have employed {\tt combine='spw'} for more S/N (and it would also be important to set {\tt interp='linearperobs'} when applying such a table to prevent interpolation across executions for this case). 

Examples of the phase vs time solutions resulting from the first iteration of self-cal are shown in Figure~\ref{MiraContPhase}. No solutions failed and the solutions are smoothly varying (as opposed to completely random), and thus are plausibly tracing true variations. Upon application of these solutions the image significantly improves with now $S/N=598$ (Fig~\ref{MiraContself}b). In this image the reality of Mira~B is confirmed (it was barely detected in the first image) and is included in the clean mask. For the second round of self-cal, $t_{solint}$ was decreased to 12~s ($4\times$ the visibility integration time). Representative phase vs time solutions are shown in Fig.~\ref{MiraContPhase} and the improvement in the image is shown in Fig~\ref{MiraContself}c. From the phase solution plots, we can see the detailed behavior of the phase variations, and it is clear that any remaining variations on timescales of 3-12 seconds (i.e. the visibility integration time up to the current $t_{\rm solint}$) will be quite small (less than a few degrees). Corrections of this magnitude can only affect the image results for dynamic ranges $\gtrsim$ a few thousand, so we conclude that we have taken the phase-only portion of the self-cal as far as it can go.  Indeed, we find that pushing the {\tt solint} down to the visibility integration time of 3~s (most easily accomplished by setting {\tt solint='int'}) shows very little additional improvement. Next we performed an amplitude self-cal while applying the second phase-only ({\tt solint='12s'}) self-cal on-the-fly. The {\tt solint} was set to 300~s since we expect it to be more slowly varying than phase and {\tt gaintype='T'}. The solutions for all antennas are plotted in Figure~\ref{MiraContAmp} demonstrating that all of the corrections are relatively small, however, as shown in Fig~\ref{MiraContself}d there is again a significant improvement with the S/N rising to 1130. It is always recommended to check the uv-data as well as the images when doing self-cal, especially after any amplitude calibration. Since amplitude calibration is less constrained, especially if the data is near the threshold for viable self-cal, it is not uncommon to get some spurious solutions that show up as bad or discrepant points in the corrected uv-data. In that case one can flag the discrepant data (if only a few points) or change the parameters of the solve to avoid getting spurious solutions. Figure~\ref{MiraContUV} shows uv-plots of the amplitude versus baseline length measured in wavelengths for several of the steps in the self-cal process. Note that for this case, the amplitude values did not change very much, which is consistent with the small deviations from 1 in the amplitude solutions. This is a good indication that we are not missing significant flux in the model.

As a final check, for the Mira continuum example above, the position of the peak pixel remains unchanged throughout the self-cal process (these images have 4~mas pixels). Examining the results more deeply, Gaussian fits to Mira~A for the original continuum image versus the final self-calibrated image reveal that differences in the peak fitted positions are less than 0.1~mas, i.e. less than the statistical errors associated with the fitting itself. Even so, when quoting relative position uncertainties, the most realistic estimate is the (synthesized beam width)/($2\times$S/N) in the original image, rather than the self-calibrated one. Regarding the absolute position uncertainty, achieving values better than $0.05''$ generally requires observation of multiple calibrators in order to assess the systematic phase error between target and calibrators \citep{Fomalont99}.

\subsubsection{ALMA Mira Spectral Line}
\label{MiraLine}

Next we discuss how the procedure needs to be modified in order to use the strongest Mira spectral line (masing \SiO) instead of the continuum. The key for this type of self-cal is to limit the solve steps to just channels with strong  emission. However, figuring this out can be tricky because most scientifically useful image cubes are made taking into account the Doppler shift due to the Earth's motion, in a time-stationary frame like the Local Standard of Rest (LSRK) or with respect to the Solar Barycenter (BARY), while the data are taken at a fixed sky frequency in the Topocentric (Earth-centric, or TOPO) frame. Furthermore, image cubes are often made averaging over several channels to increase S/N, and over only a portion of the full spectral window. These facts mean that there is no one-to-one correspondence between the input uv-data channels and the channels of the output image cube. In CASA, when you make an image cube with a time-independent frame, it regrids the data to the desired output frame but also ``de-grids'' the cube image model back to the original input data channels. So using for example {\tt plotms} it is quite easy to find out what input channel(s) contain the strongest cleaned emission by plotting the model as a function of channel. 

\begin{figure}[h!]   
\centering
\includegraphics[width=0.45\textwidth]{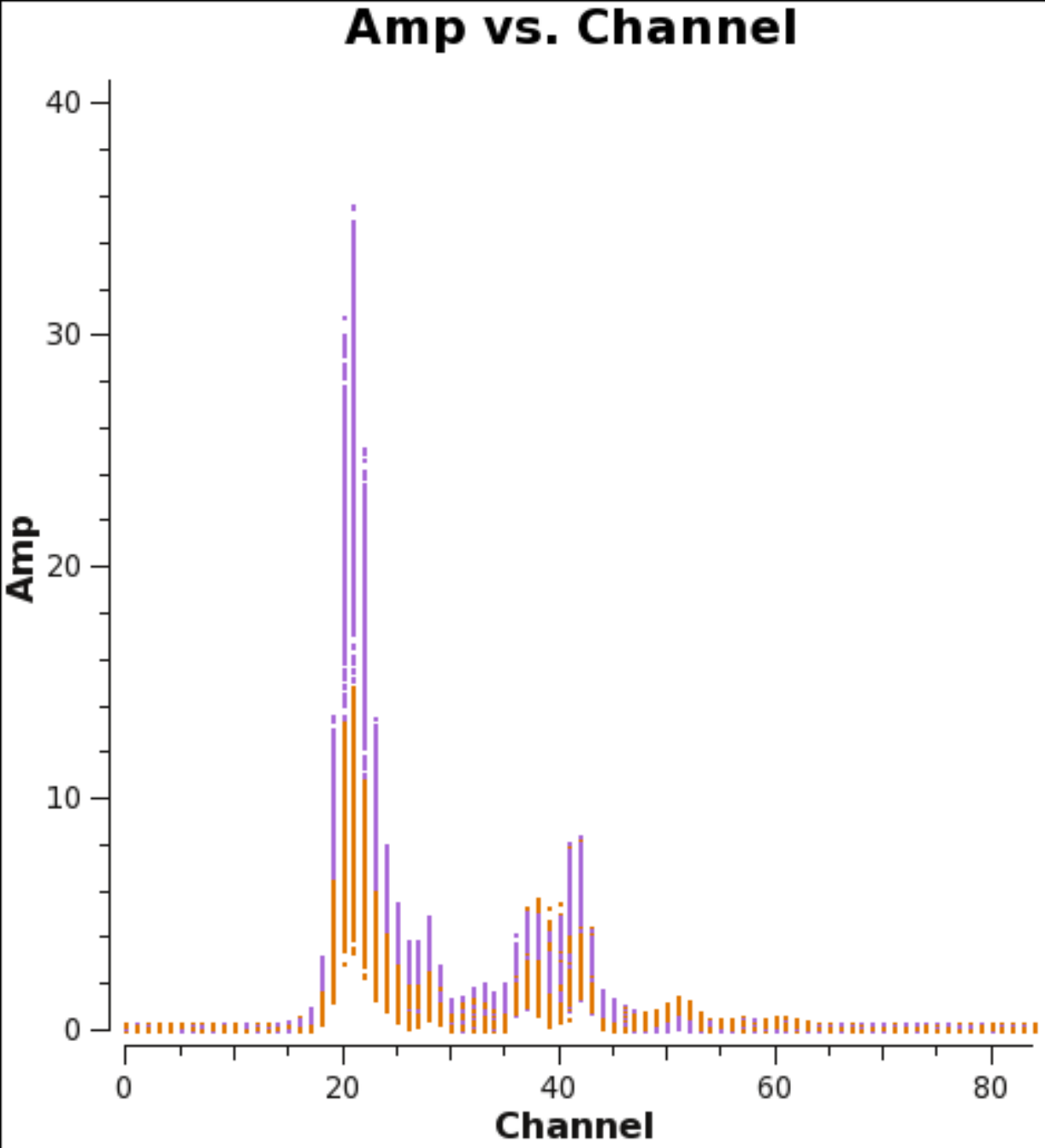}
\includegraphics[width=0.45\textwidth]{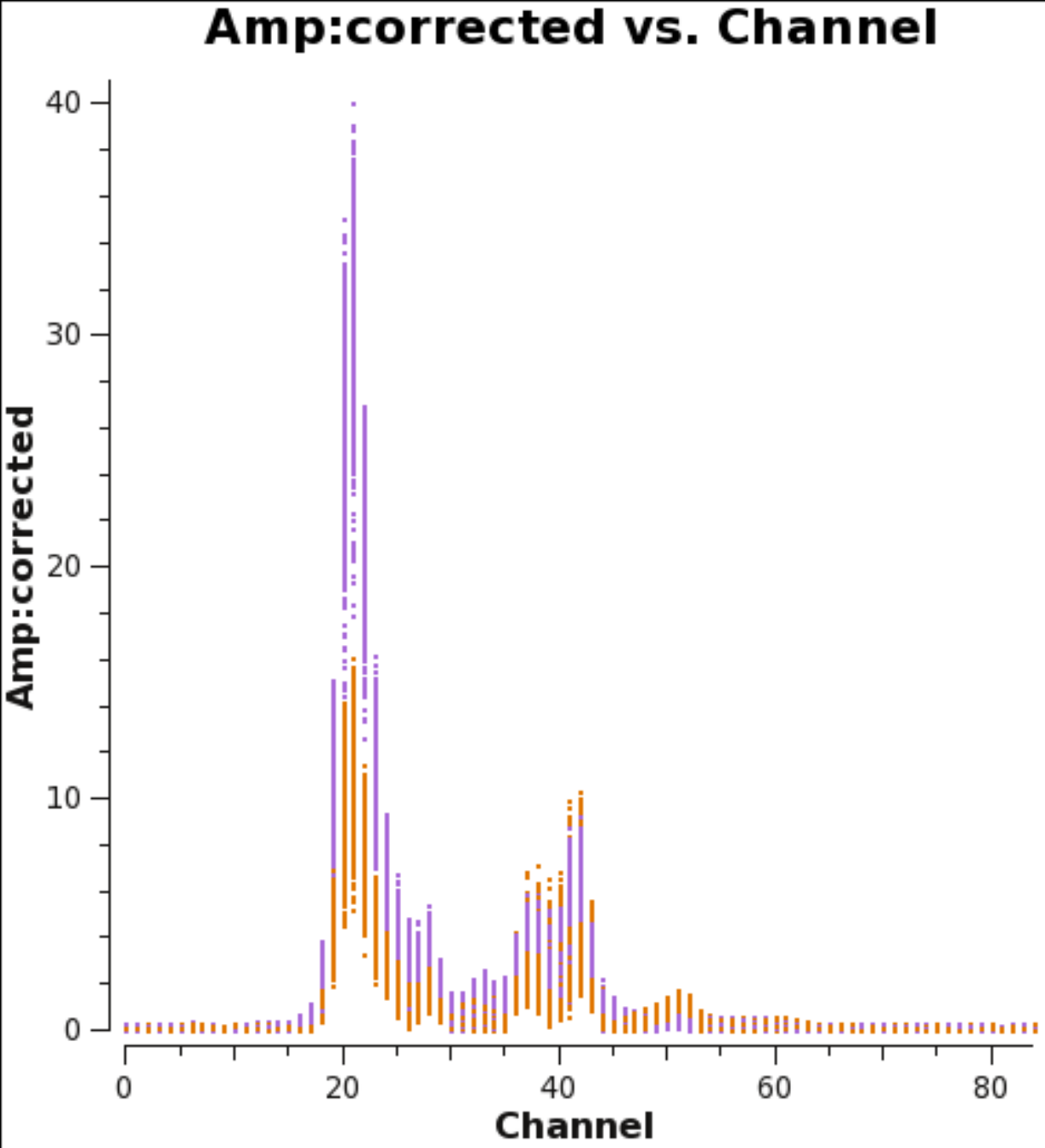}
\caption{UV spectra of the Mira \SiO\/ data.  LEFT: original calibration; RIGHT: after phase and amplitude self-calibration (points are colored by polarization). Note the increase in amplitude in all channels with emission.}
\label{MiraSioSpectra}
\end{figure}

\begin{figure}[h!]   
\centering
\includegraphics[width=0.48\textwidth]{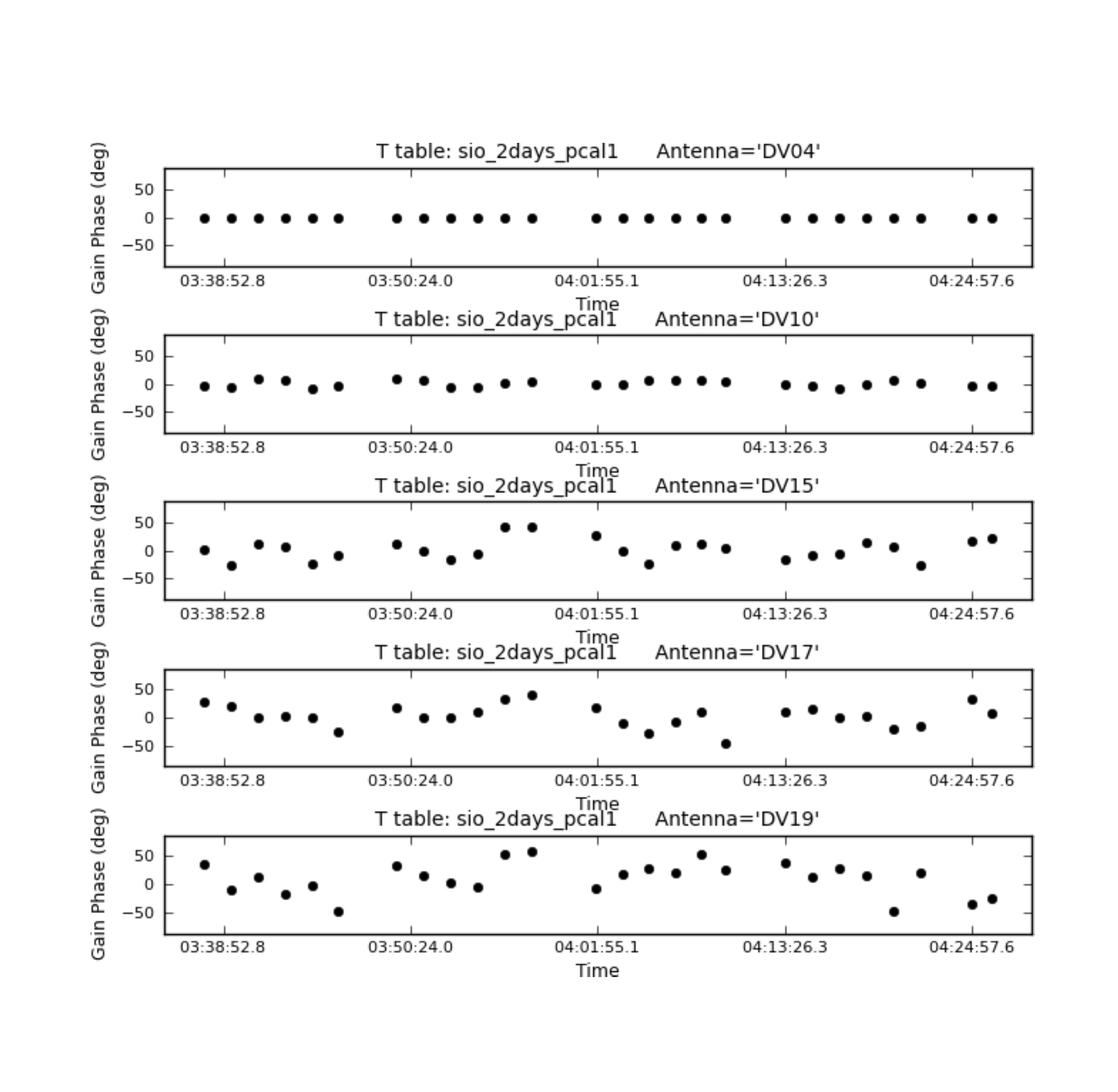}
\includegraphics[width=0.48\textwidth]{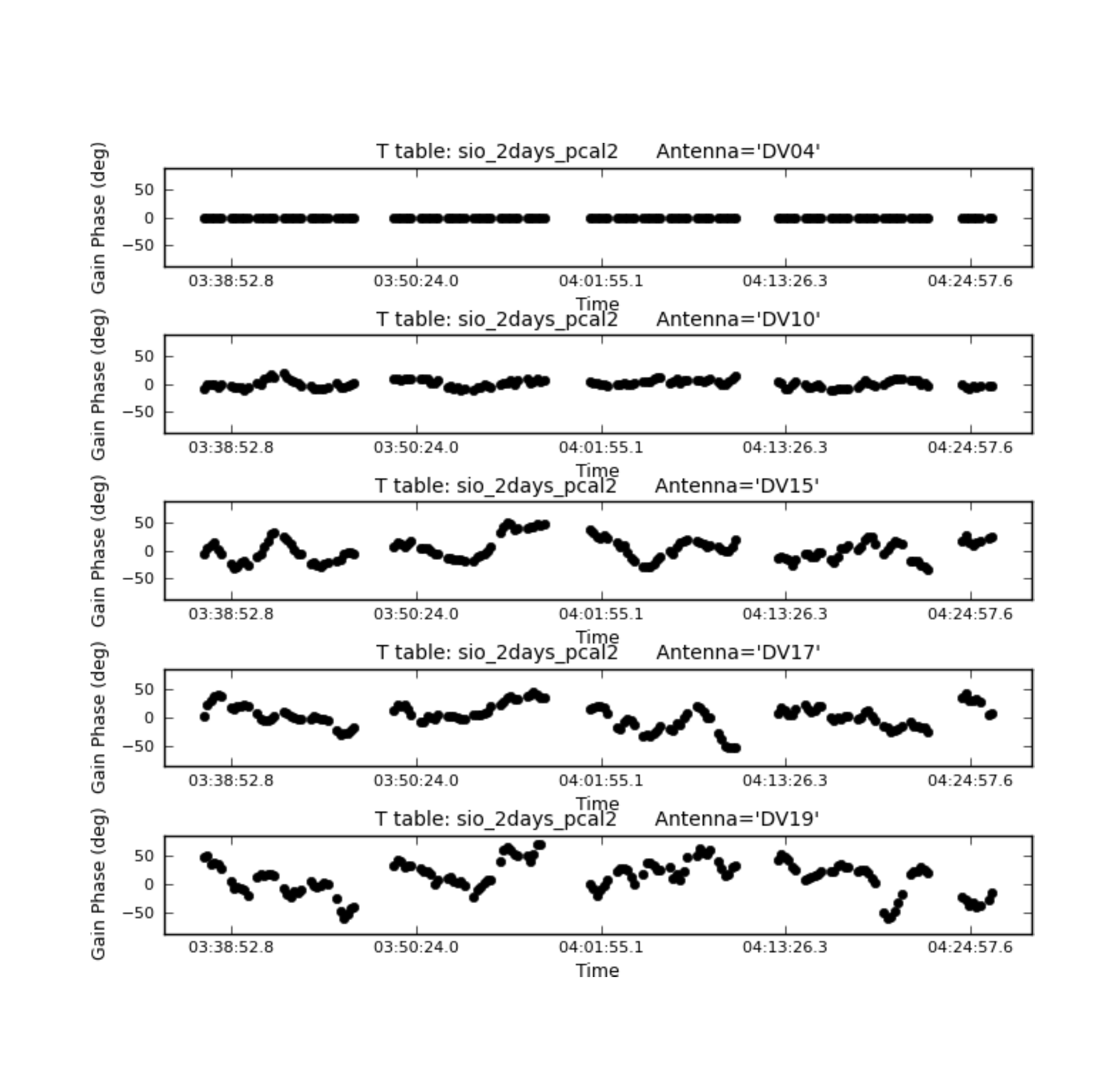}
\caption{Representative phase vs. time plots of the Mira \SiO\/ phase-only self-cal solutions for the second execution. LEFT: First phase-only iteration with {\tt solint}=scan length. RIGHT: Second phase-only iteration with {\tt solint=12}~seconds. The plotted antennas are ordered with the reference antenna in the top row and antennas located increasingly further from the array center in subsequent rows.}
\label{MiraSioPhase}
\end{figure}

\begin{figure}[h!]   
\centering
\includegraphics[width=0.85\textwidth]{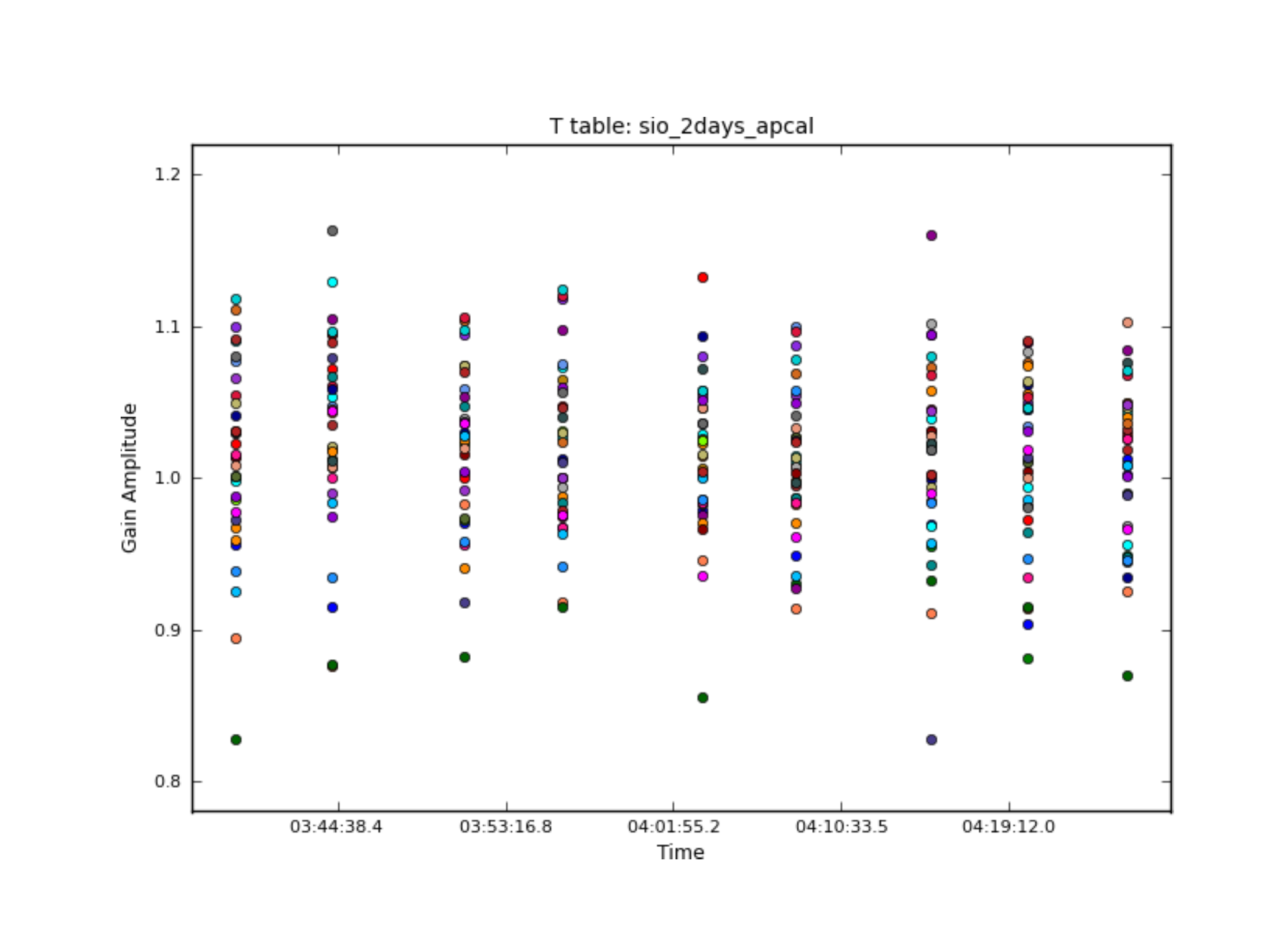}
\caption{Amplitude vs. time for Mira \SiO\/ amplitude self-cal solutions for the second execution (points are colored by antenna). The second phase-only solution was applied on-the-fly before solving for these solutions.}
\label{MiraSioAmp}
\end{figure}

\begin{figure}[h!]   
\centering
\includegraphics[width=1.0\textwidth]{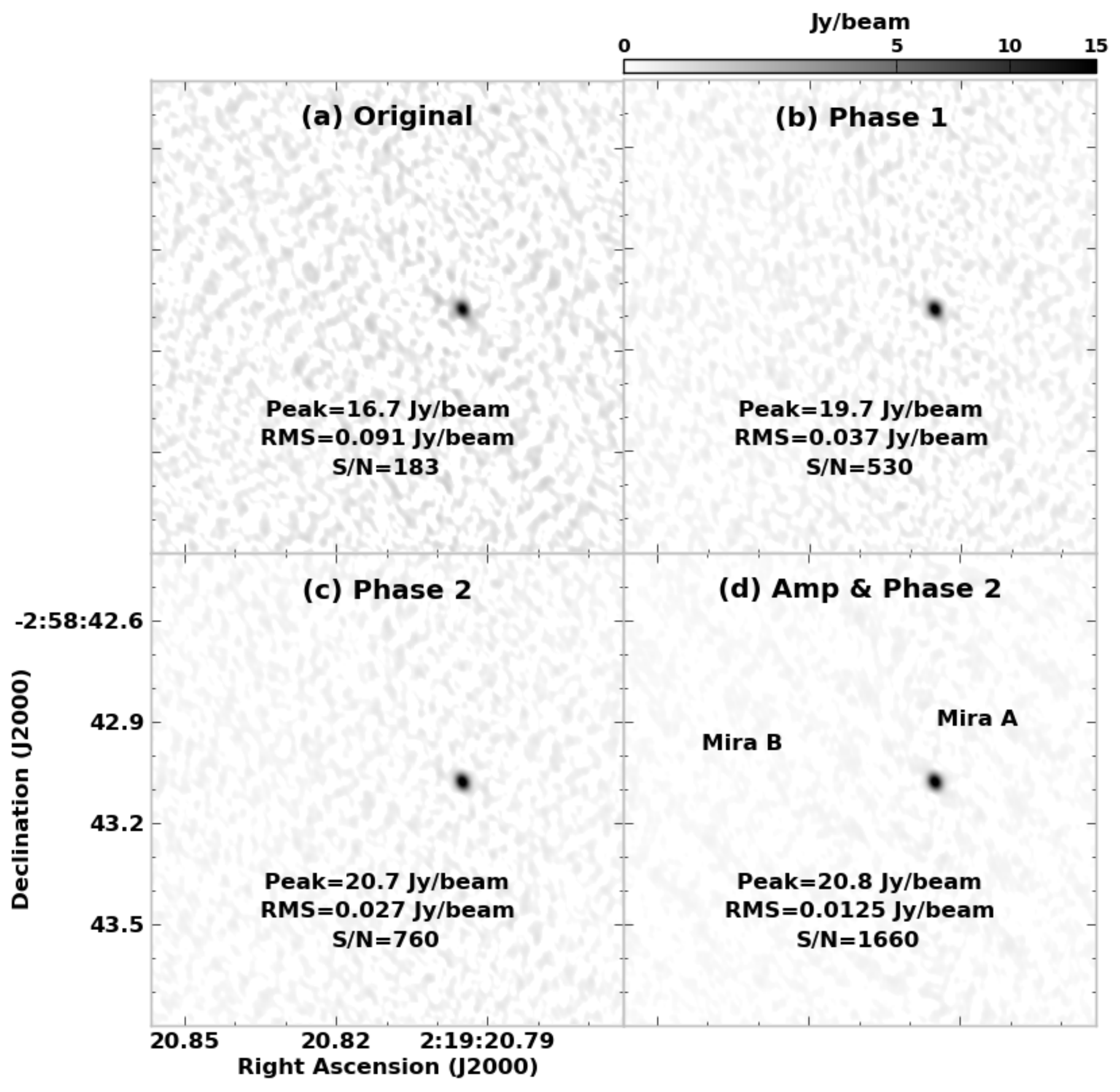}
\caption{Panels show sequence of images with indicated self-calibration for ALMA \SiO\/ spectral line data at 1.3~mm for the evolved star Mira. Note that the panels show the same log greyscale.}
\label{MiraLineself}
\end{figure}

However there is often another issue stemming from the fact that many modern radio telescopes build up the total time needed by observing projects in short executions of $\sim$1~hr duration, and use a technique called ``Doppler Setting'' to adjust the observed sky frequency just before observing begins to account for the (changing) source Doppler shift in order to keep lines centered where the user has requested them to be. This means that no two executions have exactly the same sky frequency, so that every execution that is concatenated together adds to the number of distinct spectral windows in the combined uv-dataset. The TOPO frequency offsets can be (equivalent to) up to $\pm$30~\kms\/, or hundreds of channels for high spectral resolution data taken several months apart. It can be a painful bookkeeping problem to figure out what channel the peak line emission corresponds to in each dataset. In CASA, this problem is easy to overcome by running the task {\tt cvel} on the spectral line data (typically after continuum subtraction) using the exact output cube specifications that you would like for your image cubes (channel width, start, number of channels, output frame, and line rest frequency). This task will create a new uv-data set gridded to the specified parameters, combining overlapping spws, in exactly the same way that the imaging task would. After this step, there is an exact 1-to-1 correspondence between the new cvel uv-dataset and the output cube, provided you do make the image cube with the same parameters as used in {\tt cvel}. 

To perform the Mira spectral line self-cal we have first subtracted the continuum using line-free channels in the uv-plane\footnote{It is not strictly necessary to subtract the continuum before doing the line self-cal, the combined line and continuum signal can also be used, provided the continuum adds extra signal at the spectral resolution chosen for the self-cal of the "line" emission (not the case for this Mira example).}  and then run the {\tt cvel} task to grid the \SiO\/ data from the two executions together into one spw in the LSRK frame. This line shows strong maser emission toward Mira~A (see uv-plot before self-cal in Figure~\ref{MiraSioSpectra} (LEFT)), and since the emission is very compact, it makes an excellent self-cal target.  These data do have an additional complicating wrinkle: this maser transition is significantly linearly polarized (see Fig.~\ref{MiraSioSpectra}), though no polarization calibration data were taken. It is also notable that the level of polarization changes across the maser spectrum. In such a case, one should definitely use {\tt gaintype='T'} for the phase and amplitude solves.  Although it will not give the best possible results for the single channel chosen for the self-cal, it is necessary in order to transfer the solutions across the spectrum. This is because the XX/YY amplitude ratio changes with time due to rotation of the parallactic angle (see \LectureEE\/), while their average should remain comparatively constant. If one used {\tt gaintype='G'} to solve for the XX and YY amplitude solutions independently, then this spurious change in the ratio of XX to YY would be transferred to the other channels, whose emission may have an inherently different ratio. From the Fig.~\ref{MiraSioSpectra} (LEFT) uv-plot of the \SiO\/ line before self-cal, the strongest maser channel was identified (21) and used in the {\tt gaincal} solve steps. Besides the spectral line preparation steps and the use of a single channel, the line self-cal was performed the same as the continuum self-cal described above. Figures~\ref{MiraSioPhase} and \ref{MiraSioAmp} show the solutions from the two phase iterations, and the single amplitude iteration, respectively. The resulting series of images is shown in Figure~\ref{MiraLineself} along with the image peak, rms, and S/N values for each step. The level of improvement is greater than that achieved for the continuum data, likely due to the absence of delay-like errors in the narrow channel. The final calibrated spectrum is shown in Fig.~\ref{MiraSioSpectra}.  Note that when applying these single channel self-calibration solutions to the rest of the \SiO\/ channels, one should use the {\tt interp='linearPD'} option in order to account for the expected linear change in phase as a function of frequency. 

For illustrative purposes we also performed the complete self-cal sequence for the strongest \SiO\/ maser channel two additional ways: (1) using {\tt solmode='G'} for each of the solve steps (i.e. XX and YY solutions solved for independently), and (2) same as (1) but with the additional modification of creating Stokes 'IQ' images (and thus models) at each stage, instead of only Stokes 'I', i.e. cleaning both the I and Q polarization images independently. From these two permutations, we find final S/N ratios of 3756 and 4072, respectively (both more than a factor of two better than the 'T' solution S/N of 1667). As expected, the 'G' type solutions produce a significantly better result because the strong linear polarization signal has been accounted for, and additionally, also using a Stokes='IQ' model yields the highest S/N Stokes I image for this channel. Most of the significant difference between the 'G' and 'T' solves comes from the amplitude self-cal step.  

Next we imaged one of the other strong maser channels (channel 40) located well away from the strongest channel (21, also see Fig.~\ref{MiraSioSpectra}) with no self-cal applied. Then this channel was imaged after applying the self-cal solutions obtained from all three methods employed for the \SiO\/ strongest maser channel. For the no self-cal, 'T' solutions, 'G' solutions, and 'G' solutions with Stokes='IQ' solutions methods, the channel 40 S/N ratios are 188, 395, 226, and 381, respectively. As expected, the 'T' solutions yield the best result when transferred across the variably-polarized spectrum.  It is also notable in contrast that applying the continuum self-cal solutions to channel 40 only yields a S/N=213, better than no self-cal, but not nearly as good as obtained from applying the strong \SiO\/ channel 'T'  solutions. Similarly, when the continuum self-cal solutions are applied to the strongest \SiO\/ maser channel (21), the S/N=249, again better than no-self cal (183) but significantly worse than any of the other permutations that were solved on itself.


\subsubsection{ALMA Mosaic on 30 Doradus}

\begin{figure}[h!]   
\centering
\includegraphics[width=0.85\textwidth]{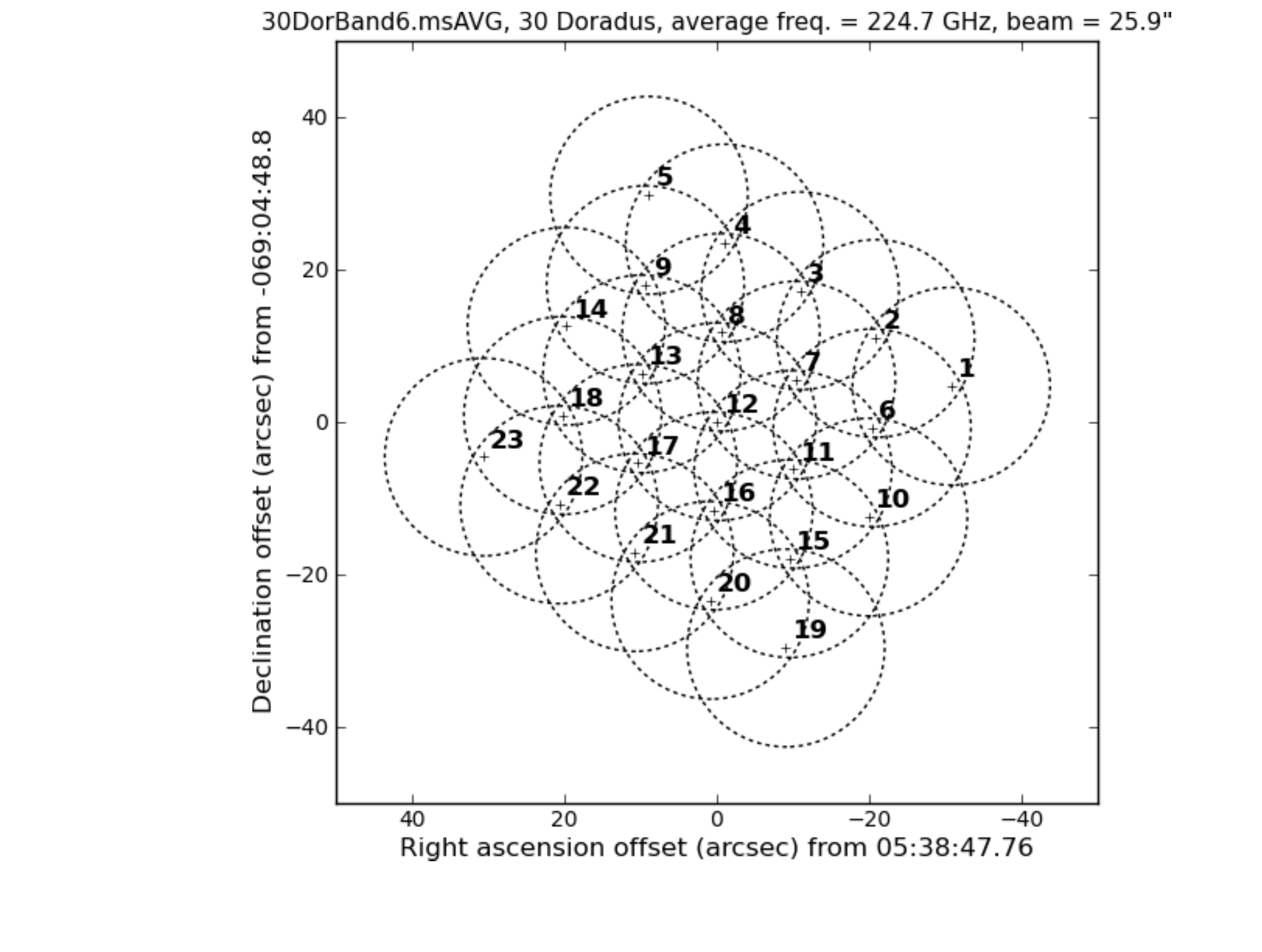}
\includegraphics[width=1.0\textwidth]{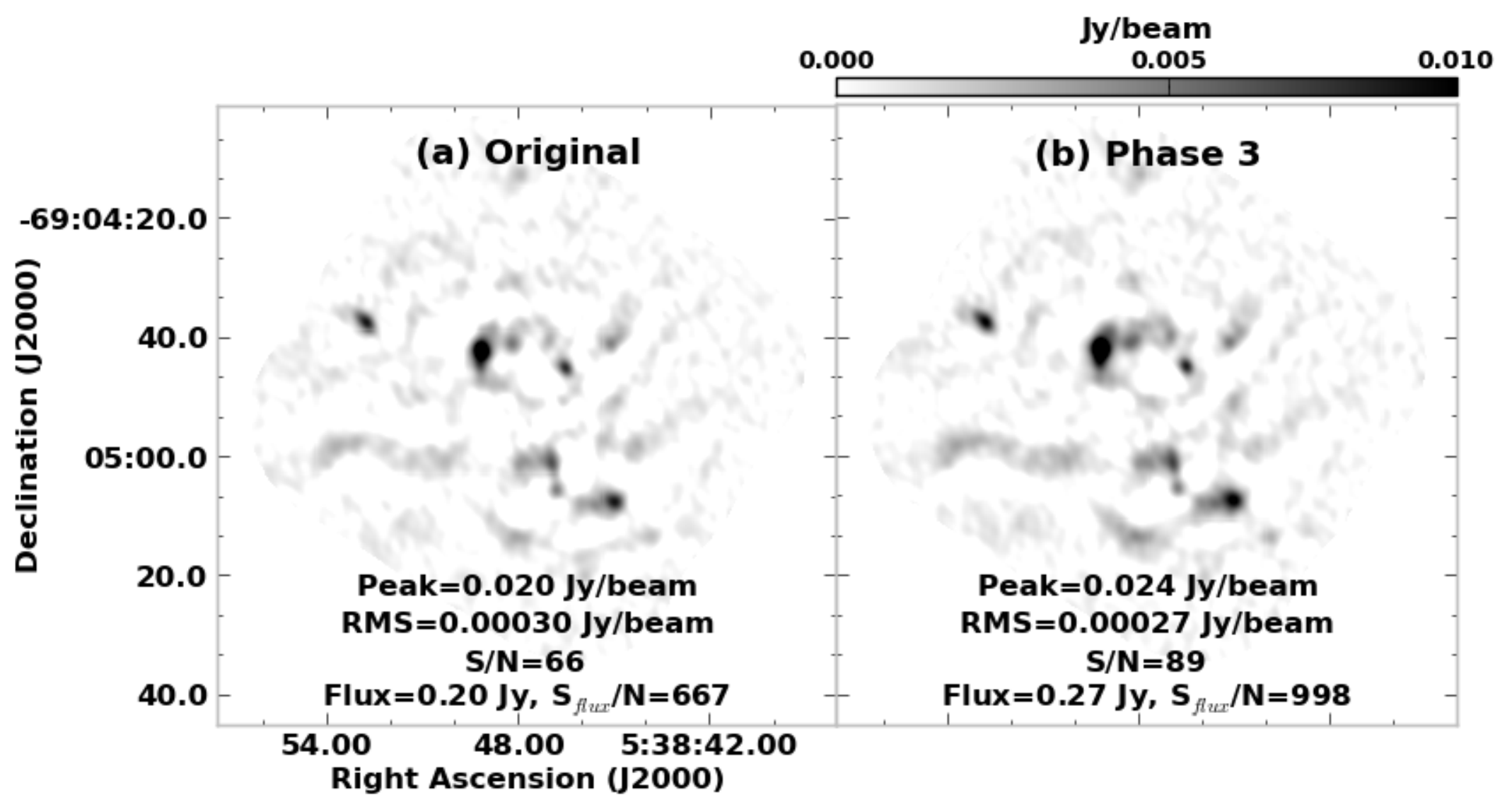}
\caption{TOP: the layout of the 23-pointing ALMA Band 6 mosaic on 30~Doradus \citep{Indebetouw13}. The field IDs are indicated for each pointing. BOTTOM (a) Original image prior to self-cal.  BOTTOM (b) Image after third iteration of self-cal. The bottom two panels are displayed on the same logarithmic greyscale. The integrated flux densities reported on the bottom panels correspond to the emission within the final clean mask.}
\label{30Dor_pointings}
\end{figure}

The final self-calibration example is presented for Cycle 0 ALMA 1.3~mm (Band 6) continuum observations toward a molecular cloud in the northern part of 30~Doradus, a young super star cluster in the Large Magellanic Cloud \citep{Indebetouw13}. The data consist of a 23-pointing mosaic (see Figure~\ref{30Dor_pointings}) observed in each of four executions. These data were obtained in a much more compact configuration (max baseline length 210~m, resolution $2.43\arcsec\times 1.82\arcsec$) than the Mira example, and include only about half as many antennas (12). For the purpose of self-cal, it is useful to know that the way that ALMA observes mosaics is to place the subset of pointings observed between two successive observations of the phase calibrator into the same ``scan'' (though each pointing will have an independent field identification number). For this project, the time between phase calibration scans is about 11.2 minutes, so that each pointing (per subset) is only observed for about 30~seconds per visit. Unfortunately, this means that only a limited amount of time averaging can be applied to the solutions without incurring a considerable time-gap  -- i.e. a relatively long time passes before subsequent observations of the same field (about 16 minutes in this case). Thus, for mosaics it can be useful to combine the data for fields within the same scan to achieve better S/N. Another critical factor in mosaic self-cal is to only include fields that contain significant emission in the solve.

\begin{figure}[h!]   
\centering
\includegraphics[width=0.9\textwidth]{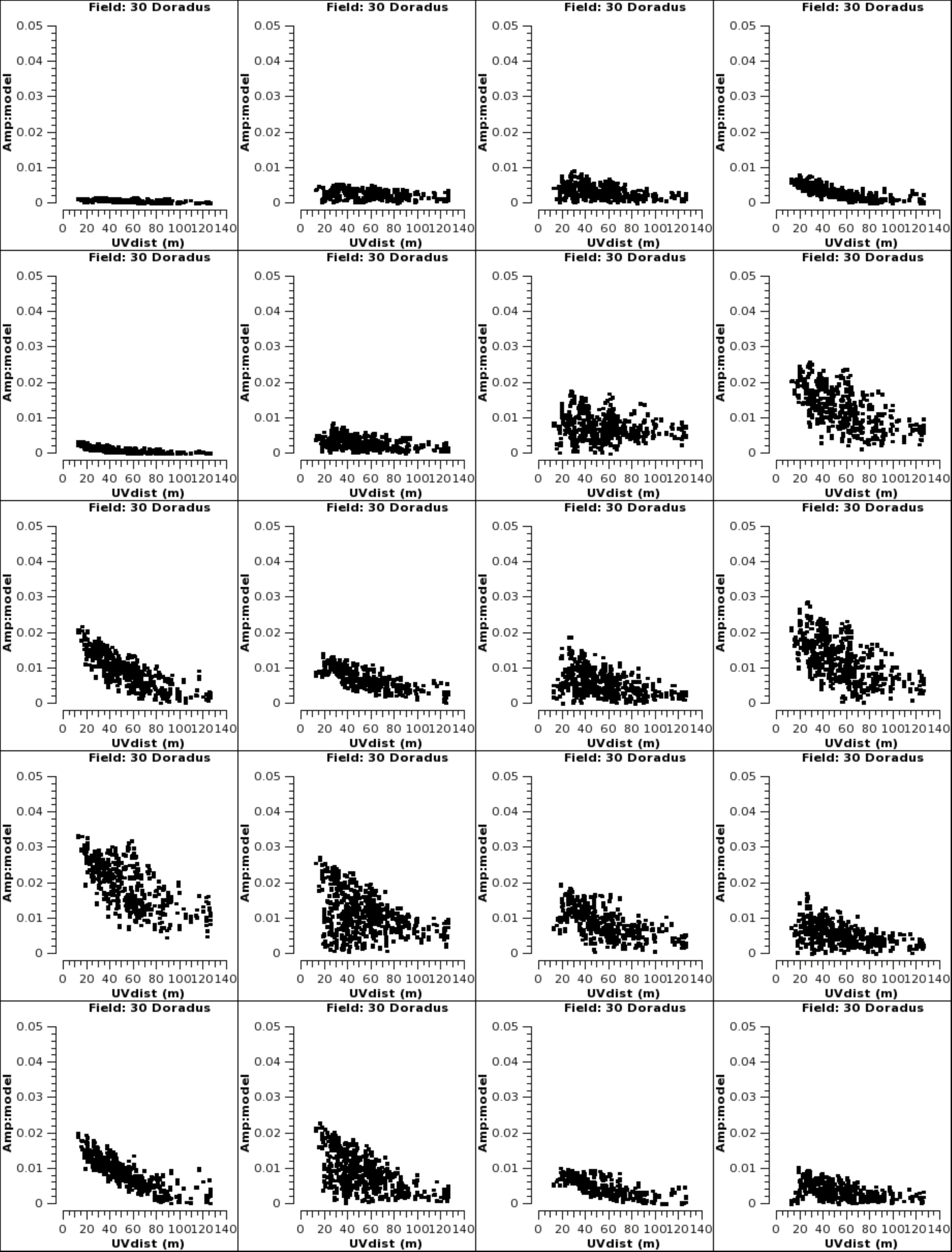}
\caption{Panels show amplitude vs uv distance of the clean model for 20 of the 23 pointings of the 30~Doradus mosaic prior to self-cal.  Fields not shown are 0, 21 and 22. }
\label{30Dor_uvamp_init}
\end{figure}

\begin{figure}[h!]   
\centering
\includegraphics[width=0.48\textwidth]{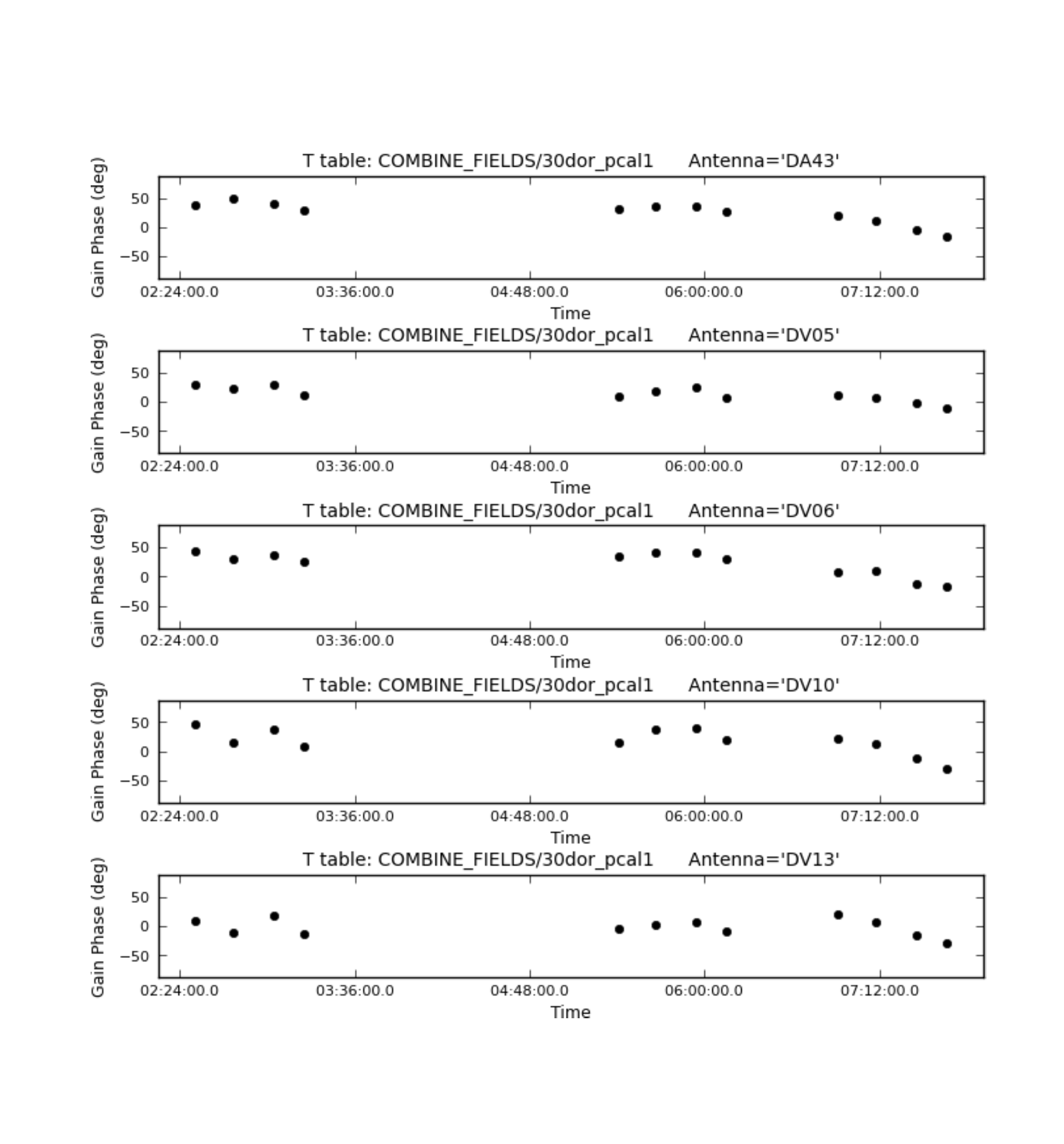}
\includegraphics[width=0.48\textwidth]{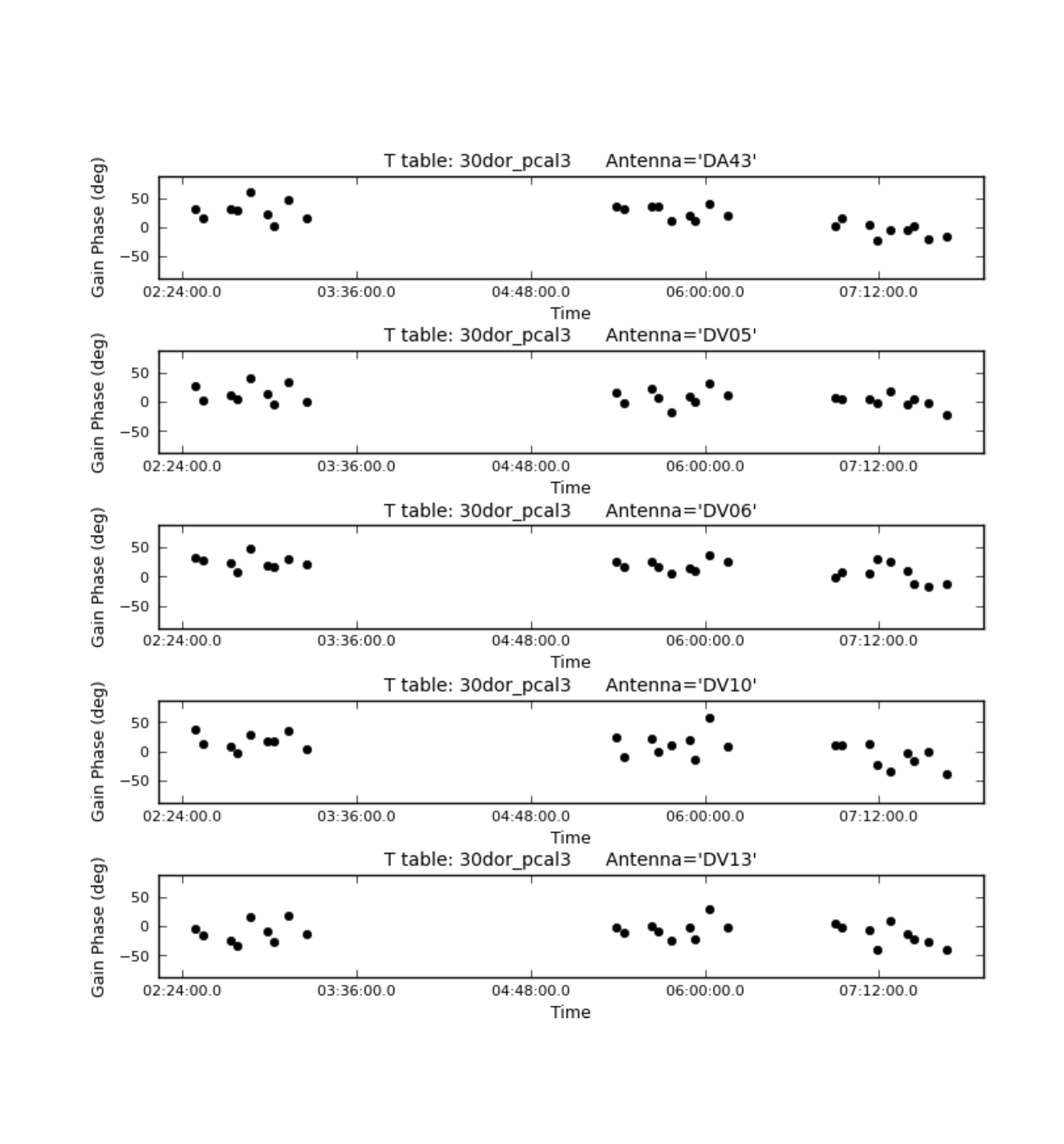}
\caption{Representative phase vs. time plots of the 30~Doradus phase-only self-cal solutions for the first three executions. LEFT: First phase-only iteration with $t_{\rm solint}$=scan length. RIGHT: Third phase-only iteration with $t_{\rm solint}$=190 seconds. The plotted antennas are ordered with antennas located increasingly further from the array center in subsequent rows.}
\label{30DorPhase}
\end{figure}

The bottom left panel of Fig.~\ref{30Dor_pointings} shows the original mosaic image, along with the starting peak, rms noise, peak S/N, integrated flux density, and the corresponding integrated flux density/rms noise (S$_{flux}$/N) in the image. The more extended nature of the emission combined with only half the antennas of the Mira examples, make these data significantly more challenging to self-cal. It is notable that when compared against the layout of the mosaic pointings, only a few of the fields have relatively strong emission. The detailed distribution of flux density per pointing in the model can be checked using {\tt plotms}, see Figure~\ref{30Dor_uvamp_init}.  Some fields show virtually no emission (first panel), while a few show significant emission in the clean model (panels 12 \& 13). As an initial trial, we attempted a phase-only self-cal for  the strongest field 13, with {\tt gaintype='T'}, {\tt combine='spw'}, and {\tt solint='inf'} to obtain one solution for this field for each of the four scans per execution. We found that even for this strongest field there were a significant number of failed solutions. From this trial, it was clear that more S$_{self}$/N would be required to self-cal these data.

\begin{figure}[h!]   
\centering
\includegraphics[width=0.9\textwidth]{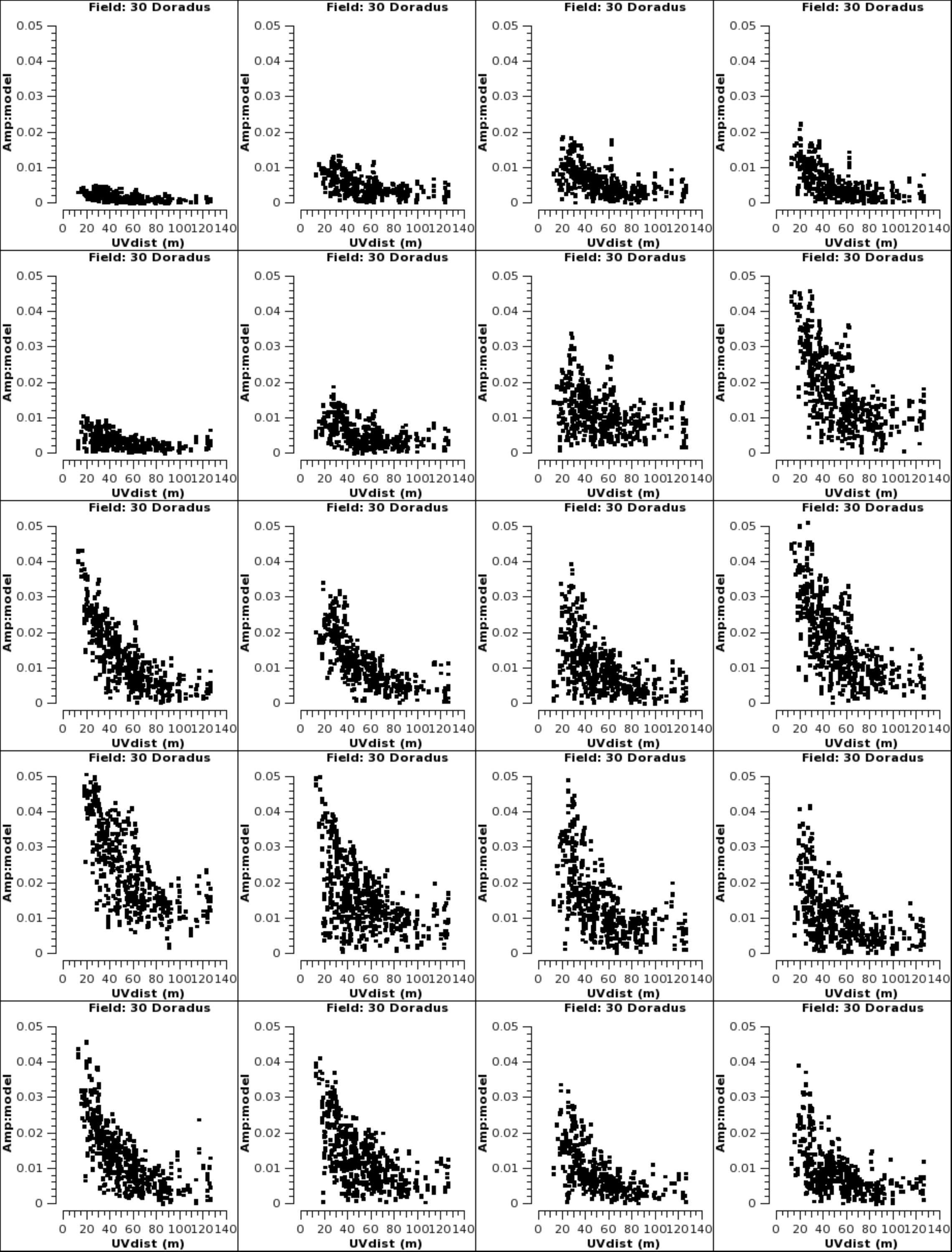}
\caption{Panels show amplitude vs uv distance of the clean model for 20 of the 23 pointings of the 30~Doradus mosaic after applying the third phase-only self-cal iteration. Note the increase in amplitude compared to Figure~\ref{30Dor_uvamp_init}.}
\label{30Dor_uvamp_p3}
\end{figure}

Next, using the Fig.~\ref{30Dor_uvamp_init} plot as a guide, we started (a new) first phase-only iteration of self-cal using the seven strongest fields (8, 9, 12, 13, 14, 17, \& 18) in the solve, along with {\tt gaintype='T'}, {\tt combine='spw,field'}, and {\tt solint='inf'}. Four solutions were obtained per execution, per antenna, and after combining the seven strongest fields, no solutions failed. Because there are multiple executions and we used {\tt combine='spw'}, it is important to set {\tt interp='linearperobs'} in the {\tt applycal}. The level of peak S/N and S$_{flux}$/N improvement after the first iteration is a factor of 1.28 and 1.34, respectively. For the second self-cal iteration, the {\tt solint} was reduced to 300 seconds and {\tt minsnr=2} instead of 3 was employed (the latter change was attempted to reduce the number of failed solutions). Additionally, based on a new uv-plot of the model, the number of fields included in the solve expanded to 12 ({\tt field='7$\sim$18'}). Of the 288 solutions (24 intervals $\times$ 12 antennas), only 1 solution failed and applying them yielded a gain of a factor of 1.28 in peak S/N and 1.37 in S$_{flux}$/N compared to the original, though this represents little change compared to the first iteration. For the third iteration, the only change (compared to the second), was to decrease the solint further to {\tt solint='190s'} and widen the field selection to 14 fields ({\tt field='7$\sim$20'}). For this iteration there were 36 $\times 12$ = 432 solutions attempted and 10 solutions failed (Figure~\ref{30DorPhase}). The improvement (over the original) in peak S/N is a factor of 1.35 and in S$_{flux}$/N is a factor of 1.50; the new clean model per pointing is shown in Fig.~\ref{30Dor_uvamp_p3}. The increasing number of failed solutions with decreasing $t_{\rm solint}$ is a signal that any further reduction in $t_{\rm solint}$ is unlikely to bring additional improvement.  Indeed, a check of the S/N of the solutions shows that many of the solutions are near the {\tt minsnr=2} limit. 

We then performed several tests of amplitude self-cal, while applying the third phase-only solutions on-the-fly. First we did a {\tt calmode='ap'} solve with a {\tt uvrange='>35m'} and one solution per execution per antenna ({\tt solint='inf', combine='spw,field,scan'}). This test yields a substantial drop in the peak and integrated flux densities despite the uv-range cutoff. Attempts to use a more restrictive uv-range cutoff led to significant data flagging. Thus, these data cannot be improved with this approach. Next we tried {\tt solnorm=True} instead of the {\tt uvrange='>35m'}, which yielded almost no change (nor improvement) in the peak intensity, rms noise, and integrated flux density. Next, $t_{\rm solint}$ was also reduced to yield one solution per scan per execution, but this also did not produce any additional improvement. A check of the S/N of the self-cal solutions reveals that many antennas are near the S/N=10 level, suggesting that further lowering of $t_{\rm solint}$ will result in unacceptably large errors in the solutions. Indeed, with the modest improvement achieved from the phase-only self-cal, it would be unusual to see a significant improvement with amplitude self-cal.

Thus, from these self-cal tests, we choose the image created from the third phase-only self-cal iteration as the final image. Over the self-cal process there is an $11\%$ improvement in the rms noise level, a $33\%$ improvement in the peak S/N, and a $50\%$ improvement in the S$_{flux}$/N (see final image in right, bottom panel of Fig.\ref{30Dor_pointings}). Though a bit trickier than the Mira examples, which had only compact emission, the self-cal result for the more extended and weak 30~Doradus 1.3~mm continuum data still yields a significant improvement, particularly in the integrated flux density, that is well worth the effort.

\section{Phase Correction Techniques}

As we have discussed, self-calibration is a phase (and amplitude) correction technique that happens offline, after observations are taken.  Unfortunately, not every science target is bright enough to self-cal.  In this situation, the best you can do is to design an observing strategy that promotes more accurate calibration and try to ensure that observations are acquired in suitable conditions.  Several observational techniques have been developed that deliver improved phase correction, and many of these are summarized in an Astro 2010 white paper \citep{Woody09}.  Here we review the nature of phase fluctuations and discuss some practical details of the correction techniques currently in use and under development.

\subsection{Nature of phase fluctuations}

Standard calibration generally removes atmospheric variations on timescales longer than the phase calibrator cycle time, provided that the antenna positions are accurate.  To mitigate the effect of antenna position errors, it is best to minimize the angular separation between the target source and the phase calibrator \citep[see quantitative discussion in][]{Hunter16}.  However, even with a nearby calibrator, faster variations will remain in the calibrated data as a residual phase error, primarily due to turbulent fluctuations in the water vapor column above each antenna \citep{Masson94}.  The distortion of the incoming phase front can be described by an observable quantity called as the structure function, $\mathscr{D}_\phi(d)$, which is the time average of the square of the phase difference ($\phi$) between two antennas vs. the distance ($d$) between those antennas \citep{Thompson07}.  The square root of $\mathscr{D}_\phi(d)$ is often called the root phase structure function, and it generally shows a power law slope ($\alpha$).   In the Kolmogorov model of turbulence \citep{Tatarskii71} with the frozen screen assumption \citep[see, e.g.,][]{Lay97}, there is a break point where the expected $\alpha$ changes from 5/6 at short size scales (dominated by 3D turbulence) to 1/3 at intermediate size scales (dominated by 2D turbulence).  The existence of the break point, as well a confirmation of $\alpha=1/3$ at kilometer scales, was demonstrated in early VLA testing \citep{Carilli99,Sramek90}. In practice, measurements of $\sqrt{\mathscr{D}_\phi(d)}$ using dedicated 11~GHz atmospheric phase interferometers (APIs) at high altitude observatories typically show intermediate values of $\alpha \approx 0.60-0.75$ on their short baseline of $\approx300$m, including at the VLA \citep{Butler99} and SMA \citep{Hunter12,Masson94}.  The break point in the slope is sometimes called the corner point, which can be either a corner frequency or a corner time ($t_{\rm corner}$) depending on how the API data are presented.  Typical values of $t_{\rm corner}$ are 20-30~seconds at ALMA and the VLA, respectively, but it varies seasonally \citep{Butler01,Butler99}.  At typical wind speeds, $t_{\rm corner}$ corresponds to a baseline of a few hundred meters, and indeed a corresponding break point is seen in the rms phase of data from the main arrays \citep[e.g.][]{Asaki14}. At larger scales ($\gtrsim$a~few~km), the atmospheric variations will generally become uncorrelated, leading to a further break in slope to $\alpha\approx0$. This region is called the "outer scale", where the magnitude of the phase fluctuations no longer increases with baseline length.

\subsection{Fast Switching}
\label{FS}

Fast switching (FS) phase calibration is simply normal phase transfer calibration but observed with a higher cadence. This technique can remove tropospheric phase variations whose size scale is larger than an "effective" length of $\approx v_{a}t/2$ where $v_{a}$ is the atmospheric wind velocity aloft (typically 10~m~sec$^{-1}$), and $t$ is the total switching time from the beginning of one observation of the phase calibrator to the next \citep{Carilli99}. Using FS cycle times of 100-150 seconds, the effective length is kept below 500-750~m, which means that phase corrections are possible on the steep portion of $\sqrt{\mathscr{D}_\phi(d)}$, thereby limiting the damage to some degree.  After FS calibration, the residual rms phase error on baselines longer than the effective baseline will be similar to one another \citep{Holdaway92}.  FS has been demonstrated at the VLA to enable the imaging of faint sources with diffraction-limited spatial resolution on the longest baselines and highest frequencies. Growing out of that experience, the ALMA 12~m antennas were specifically designed with FS in mind with a target performance of completing a $1.5^\circ$ switch to within $3''$ error in 1.5~s \citep{Rampini10,Mangum06}.  To achieve this goal, the antennas have an angular velocity specification of $6^\circ$~s$^{-1}$ in azimuth and $3^\circ$~s$^{-1}$ in elevation, with the acceleration being 3 times those values.  The SMA antennas have a comparable performance \citep{Hunter13}.  The initial ALMA observations in the Cycle 3 long baseline configurations used FS, typically spending 15-18~s on the calibrator and 48-54~s on the science target \citep[e.g.,][]{Hunter15}.  In this mode, the current observing efficiency on the science target is about 50\%.  Somewhat longer durations are used in smaller configurations to achieve better efficiency.

\subsection{Radiometers}
\label{wvrs}
As an important complement to FS calibration, radiometric phase
correction provides a method of measuring and removing the faster
atmospheric variations (on the 1-60 second timescale) that cannot be
removed by FS alone. The definition of a radiometer is a sensitive
receiver, typically with an antenna input, that is used to measure
radiated electromagnetic power \citep{NIST08}.  In the context of an
interferometer, radiometric phase correction refers to the use of a
specialized radiometer that measures relative changes in sky
brightness temperature versus time, either in a broadband continuum
channel or in several narrow channels within an atmospheric spectral
line.  The changes in brightness temperature are converted to changes
in path length via a physical model accounting for current atmospheric
conditions and how they affect the profiles of the various water
absorption lines \citep[e.g.,][]{Sutton96}. These path changes are
subsequently converted to expected phase changes at the science
observing frequencies, which can then be applied to the raw
data. Examples of continuum-based radiometric phase correction include
the system at the IRAM Plateau de Bure Interferometer (PdBI) employed
during the 1990's \citep{Bremer96} which used the facility 1.3~mm SIS
receivers, as well as a similar tests performed at 1.3~mm at the
SMA \citep{Battat04}.

\begin{figure}[h!]   
\centering
\includegraphics[width=1.0\textwidth]{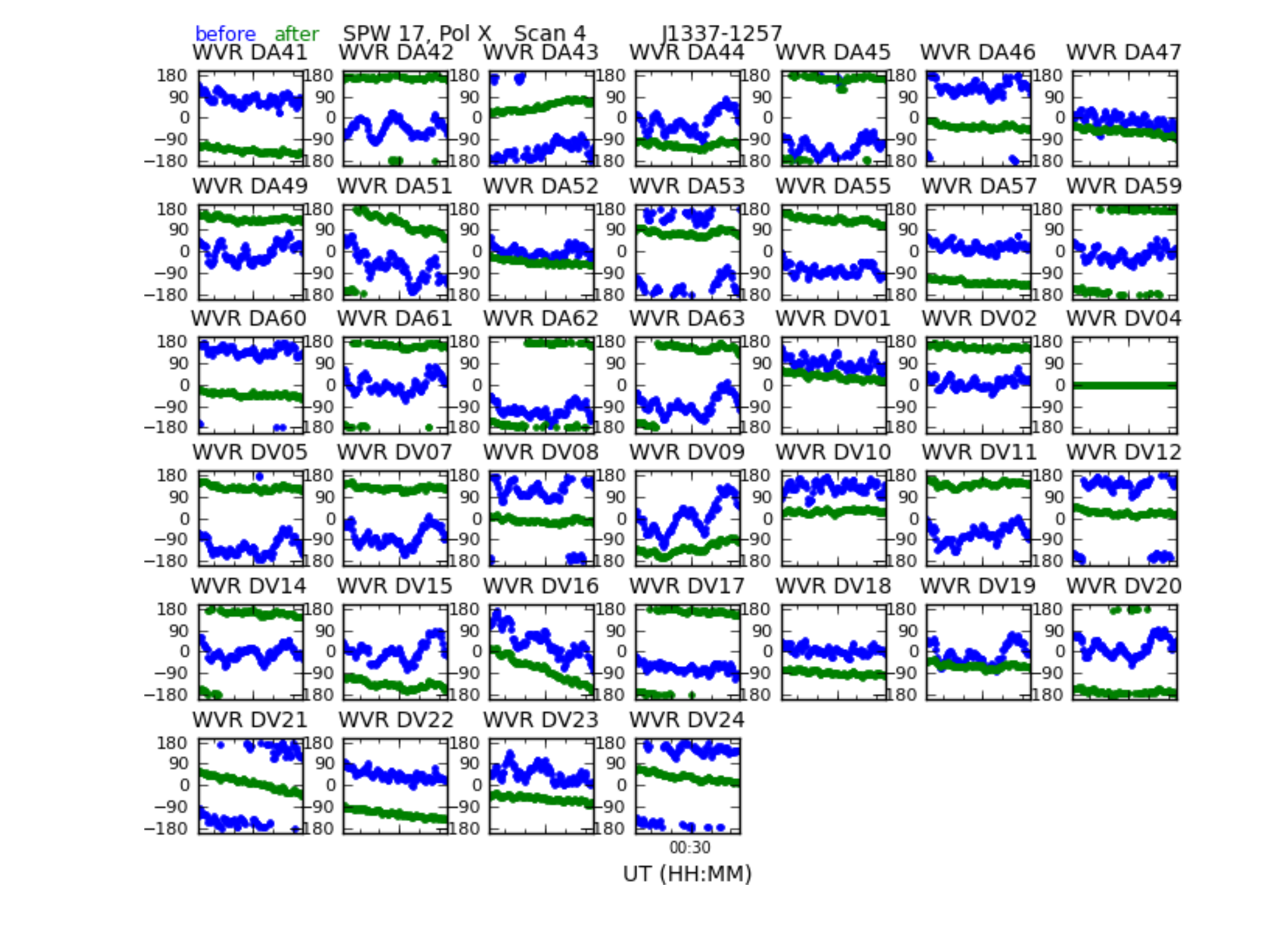}
\caption{The individual panels show the integration-based phase solutions in units of degrees vs. time for each antenna during a ten minute ALMA observation of a bandpass calibrator (J1337-1257) in Band 7. The blue points are the results for the raw data, and the green points are obtained after applying the WVR corrections from {\tt wvrgcal}. DV12 was the reference antenna.}
\label{wvrfig}
\end{figure}

The water vapor radiometers (WVRs) on each of the ALMA 12~m telescopes
provide an excellent example of a spectral radiometric phase
correction system \citep{Emrich09} that is in routine use at an
observatory.  Each WVR is a 4-channel double sideband (DSB) system
tuned to observe the wings of the strong 183~GHz water line. The WVR
data are recorded at 1.152~s intervals, which yields rms noise levels
well below 0.1~K rms in each channel.  This timescale also corresponds
approximately to the antenna diameter divided by the typical
windspeed, and thereby represents the fastest rate at which the water
column above an individual dish can change appreciably. A
noise-weighted sum of the product of the individual phase correction
coefficients (d$L$/d$T_{B,k}$) with the observed brightness
temperature fluctuations ($\delta T_{B,k}$) from each channel $k$ of
the radiometer are used along with a simple one-layer model and
Bayesian analysis to derive visibility phase corrections offline via
the CASA task {\tt wvrgcal} \citep{Nikolic13}.  These WVR phase
corrections improve the phase stability of ALMA data in most
conditions \citep{Fomalont15,Asaki14}.  An example of the improvement
on the phases obtained by applying these corrections offline is shown
in Figure~\ref{wvrfig}. Implementation of an independent algorithm to
apply phase corrections online is under experimentation in the ALMA
observatory's online calibration
software \citep[TelCal,][]{Broguiere11}, which would allow longer time
averaging of the corrected data prior to storage.

The use of WVR phase corrections are not as straightforward for
observations taken under significant clouds.  In this case, the ALMA
WVR corrections derived from individual radiometer channels (2 and 4)
are found to be quite different from one another, a condition
quantified by the term discrepancy \citep{Nikolic12}.  Applying these
solutions will degrade the raw data, evidently due to the presence of
liquid water droplets in the clouds.  These droplets will contribute
significant continuum absorption ($\tau \propto \nu^2$ when $d_{\rm
droplet}<<\lambda$), but will change the index of refraction by only a
small amount compared to water vapor \citep{Thompson07}.  Therefore,
meteorological instruments typically employ multi-channel radiometers
to assess the effect of water vapor and liquid
simultaneously \citep[e.g. 22.235, 31.65 \&
85.5~GHz,][]{Westwater80,Bobak00}. The Deep Space Network sites use
3-channel WVRs operating at 22.2, 23.8 and
31.4~GHz \citep{Morabito15}. Likewise, the water vapor radiometers
at NOEMA (in use since 2004) have three 1~GHz-wide
channels \citep[19.2, 22.0, \& 25.2~GHz,][]{Bremer06}, which
effectively discriminate between liquid water and water vapor.  In a
similar vein, fitting and removing a continuum component of the
brightness temperature due to liquid water from the ALMA WVR data
prior to solving for the water vapor path variations with {\tt wvrgcal} is
a promising technique that has shown effective results on ALMA
commissioning and science data.

\subsection{Special Phase Transfer Techniques}

When the signal to noise on the phase calibrator is high in all
spectral windows, it is usually best to determine independent solutions
per window and transfer them directly to the science target.  
However, when the signal to noise is low, one must resort to other techniques.

\subsubsection{Wide Bandwidth to Narrow Bandwidth Transfer}

Modern correlators, such as the Wideband Interferometric Digital
ARchitecture (WIDAR) correlator \citep{Perley09} and the ALMA
correlators \citep{Escoffier07,Kamazaki12}, allow one to define dozens
of spectral windows of varying bandwidth depending on the science
goal. In windows with high spectral resolution, sometimes the
bandwidth is too small to provide good S/N on the phase calibrator in
a single scan.  In this situation, one can use the time dependent
variation observed in a wide window and apply it to one (or more)
narrow windows nearby in frequency, provided that an initial
observation of a bright source has been performed in order to measure
the phase and amplitude offsets between the wide and narrow
windows. This technique has been in common use for decades at the VLA
in the form of "bandwidth switching", in which the phase calibrator is
observed in a wide band mode, and the science target is observed in a
narrow band mode (for spectral line observations). Note that the time
dependent variation being calibrated in this case is assumed to be
primarily atmospheric in origin.  If there is any time dependent
variation of instrumental phase or amplitude between the windows, then
this variation needs to be calibrated separately by periodic
observations of the bandpass calibrator in both windows.  Finally, in
high frequency ALMA observations, it is not uncommon for the S/N to be
low on the phase calibrator even in a wide spectral window that covers
all of one of the four basebands.  In this case, after determining the
spectral window offsets on the bandpass calibrator, one can use {\tt
gaincal} to combine the visibilities from all the spectral windows in
all basebands in order to obtain the most sensitive aggregate solution
on the phase calibrator.

\subsubsection{Band to Band Transfer}
\label{btob}

In the radio to submillimeter region, the atmosphere is nearly non-dispersive at all frequencies where it is transparent \citep{Liebe89}.  Thus, in principle, one
can measure atmospheric phase changes vs. time in one frequency band, scale them appropriately  to a different frequency band, and remove them from contemporaneous observations.  The two bands can be observed either sequentially, as in the case of a single-receiver system like ALMA, or simultaneously, as in the case of a dual-receiver system incorporating a wavelength diplexer, e.g. the Submillimeter Array \citep[SMA,][]{Hunter05}.  Plausible success with transferring phase using this band-to-band (B2B) technique has been demonstrated at SMA \citep{Hunter06} and in ALMA test observations, but it is so far not a common observing mode.  The transfer of amplitude variations with time (both atmospheric and instrumental) is an essential component of gain calibration which has often received less attention in this context.  We caution that compared to other techniques, B2B transfer is more reliant on instrumental stability, which must be understood and characterized at each observatory \citep[e.g.,][]{Kubo06}.  A special case of B2B transfer in millimeter VLBI has been termed "fast frequency switching" in which the same source is observed continuously but at two tuning bands interleaved in time.  The lower frequency (and thus lower angular resolution) band provides the reference for the higher frequency band to be imaged at higher angular resolution \citep{Middelberg05}.

Dispersion occurs as one approaches the strong atmospheric lines (mainly the 557, 752 and 988~GHz water lines) that define the submillimeter observing bands \citep{Sutton96}. Thus, the edges of ALMA bands 8-10 are most affected.  For example, at 630~GHz, the truth path per unit water column is 12-15\% less than the non-dispersive assumption, with the variation depending on the temperature \citep{Hunter05}.  Correcting for the effect of dispersion in bandwidth switching and B2B transfer by scaling the phase solutions (or implementing them as a delay) is a future development item for ALMA and CASA.  Observations at the ALMA band edges are more common than one might expect, primarily due to the popularity of targeted observations of high-$J$ CO and fine structure lines redshifted to non-optimal frequencies in the ALMA bands \citep[e.g.,][]{Ferkinhoff15}.

\subsubsection{Paired Array Transfer}

The concept of paired array transfer splits the available antennas into co-located pairs (within a few antenna diameters of each other) such that one element of each pair constantly observes the science target while the other element constantly observes a nearby calibrator to monitor the atmosphere \citep{Asaki96,Dravskikh79}.  The antennas must be far enough apart to avoid shadowing but close enough to maintain comparable paths through the troposphere. The gain solutions derived from the "reference" antenna are then applied to the corresponding neighboring "science" antenna.  In order for this technique to work, the instrumental gain must be stable on all antennas or at least measured periodically during the observation with the resulting corrections applied offline.  With antennas of uniform size, this technique suffers a factor of two loss in raw sensitivity on the science target because the collecting area is split in half, but it recovers a factor of $\approx\sqrt{2}$ because all of the time is spent on the target rather than $\approx 1/2$ the time in the more traditional FS approach \citep{Thompson07}.  Nevertheless, the net loss of $\sqrt{2}$ can be worthwhile as it offers more accurate phase correction, particularly during unstable conditions.  However,   
for a large array with 50 antennas, 3/4 of the baselines are lost, so imaging fidelity is heavily compromised.  To improve these efficiencies one can construct smaller reference antennas \citep[as pioneered by CARMA,][]{Perez10,Zauderer14}, or arrange the antenna configuration into triads with one reference antenna per two science antennas.  Both of these mitigating concepts are under consideration for the Next Generation VLA \citep{Clark15}.  Clearly, this technique is only practical if both the antennas and configurations are designed with it in mind.
 
\acknowledgements The National Radio Astronomy Observatory is a facility of the National Science Foundation operated under agreement by the Associated
Universities, Inc. This research made use of NASA's Astrophysics Data
System, the IEEE Xplore digital library and the SPIE digital
library. This paper makes use of the following ALMA data:
ADS/JAO.ALMA\#2011.0.00014.S and 2011.0.00471.S. ALMA is a partnership
of ESO (representing its member states), NSF (USA) and NINS (Japan),
together with NRC (Canada), NSC and ASIAA (Taiwan), and KASI (Republic
of Korea), in cooperation with the Republic of Chile. The Joint ALMA
Observatory is operated by ESO, AUI/NRAO and NAOJ. The authors thank
Catherine Vlahakis for feedback on the manuscript.



\end{document}